\documentclass[letterpaper,journal]{IEEEtran}

\usepackage{silence}
\usepackage{amsmath,amsfonts}
\usepackage{algorithmic}
\usepackage{array}
\usepackage[caption=false,font=normalsize,labelfont=sf,textfont=sf]{subfig}
\usepackage{textcomp}
\usepackage{stfloats}
\usepackage{url}
\usepackage{verbatim}
\usepackage{graphicx}
\usepackage{bm}
\usepackage{booktabs}
\usepackage{arydshln}
\usepackage{nicematrix}
\usepackage{siunitx}
\usepackage{hyperref}

\usepackage{microtype} 

\usepackage{tikz}
\usetikzlibrary{positioning, calc, arrows.meta}

\def\BibTeX{{\rm B\kern-.05em{\sc i\kern-.025em b}\kern-.08em
    T\kern-.1667em\lower.7ex\hbox{E}\kern-.125emX}}
\usepackage{balance}

\newcommand{\ssub}[1]{_{\scriptscriptstyle #1}}

\newcommand{\Ab}{\bm{A}}                              
\newcommand{\Bb}{\bm{B}}                              
\newcommand{\Abm}{\bm{A}_{-}}                         
\newcommand{\Bbm}{\bm{B}_{-}}                         
\newcommand{\Ag}{\bm{\mathsf{A}}}                     
\newcommand{\Bg}{\bm{\mathsf{B}}}                     
\newcommand{\Agm}{\bm{\mathsf{A}}_{-}}                
\newcommand{\Bgm}{\bm{\mathsf{B}}_{-}}                

\newcommand{\yb}{\bm{y}}                      
\newcommand{\xb}{\bm{x}}                      
\newcommand{\ybm}{\bm{\tilde{y}}}                      
\newcommand{\xbm}{\bm{\tilde{x}}}                      
\newcommand{\xg}{\bm{\mathsf{x}}}            
\newcommand{\yg}{\bm{\mathsf{y}}}            
\newcommand{\xbp}{\bar{\bm{x}}}               
\newcommand{\ybp}{\bar{\bm{y}}}               

\newcommand{\Hb}{\bm{H}}
\newcommand{\Hba}{\bm{H\ssub{A}}}
\newcommand{\Hbb}{\bm{H\ssub{B}}}

\newcommand{\SR}{\bm{R}}               
\newcommand{\Se}{\bm{e\ssub{0}}}               

\newcommand{\vg}{\bm{\mathsf{v}}}            
\newcommand{\wg}{\bm{\mathsf{w}}}            
\newcommand{\vbp}{\bar{\bm{v}}}               
\newcommand{\wbp}{\bar{\bm{w}}}               

\newcommand{\ye}{\bm{y_e}}
\newcommand{\we}{\bm{w_e}}

\begin{document}
\bstctlcite{IEEEexample:BSTcontrol}

\title{Fast Cascaded Recursive Filtering via a Block-Matrix Reformulation}

\author{Haotian~Zhai,~\IEEEmembership{Student~Member,~IEEE,}
        and~Bernd-Peter~Paris,~\IEEEmembership{Senior~Member,~IEEE}%
\thanks{This work was supported in part by the National Science Foundation under grant 2029836.
H.~Zhai and B.-P.~Paris are with the Department of Electrical
and Computer Engineering, George Mason University, Fairfax, VA 22030
USA (e-mail: hzhai@gmu.edu; pparis@gmu.edu).}}


\maketitle

\begin{abstract}
Recursive (IIR) filters realized as cascaded second-order
sections (biquads) offer both design generality and
robustness against coefficient quantization.  
However,
their inherent sample-to-sample feedback dependency poses
a fundamental obstacle to parallel computation.  
This
paper reformulates the biquad difference equation as a
banded block-Toeplitz linear system.
It relies on a
stride-$N$ permutation that maps a group of $NL$ samples
into a block-tridiagonal structure whose entries are
scalar multiples of identity and shift matrices.  
Within
this framework, two data-parallel algorithms are developed 
which,
to the best of our knowledge,
are the fastest, single-core IIR filter implementations reported to date.
A partial LU (PH) factorization
preserves the sparse block structure and a cyclic
reduction reduces the
sequential dependency depth from $\mathcal{O}(N)$ to
$\mathcal{O}(\log_2 N)$.
For a cascade of $K$ biquads, the intermediate
permutations between successive sections cancel exactly,
so that only a single pair of permutations is
required for the entire cascade.
Exact block-level operation counts are
derived for every algorithmic stage and validated against
cycle-accurate measurements on three 
micro-architectures supporting AVX2 SIMD instructions.
Experimental results for a 16th-order system show that the
proposed multi-block algorithms reduce clock cycles per
sample by up to $10\times$ compared to scalar filtering,
with both algorithms scaling favorably on newer
architectures.  
On a single Meteor Lake core, cyclic
reduction achieves approximately 616~MS/s---four times the
throughput of Intel's vendor-tuned IPP library.\looseness=-1
\end{abstract}


\section{Introduction}
\label{sec:intro}

\IEEEPARstart{R}{ecursive} filters are among the most fundamental
building blocks in digital signal processing.  
Compared to
non-recursive (FIR) filters, they typically require far fewer
coefficients to meet a given magnitude-response specification,
translating directly into fewer computations per sample.  This
efficiency makes them especially attractive in high-throughput
settings such as real-time digital filtering in modern communication
receivers~\cite{Valkama_01}, speech and audio
processing~\cite{Belloch_14}, high-resolution image
restoration~\cite{Nehab_11}, and video processing~\cite{Tammana_23}.
However, the inherent feedback dependency in a recursive filter, where
each output sample depends on previous outputs, creates a sequential
bottleneck that limits computational speed and has long motivated the
search for parallel alternatives.

Early efforts to break this bottleneck employed the Fast Fourier
Transform to approximate the infinite impulse
response~\cite{Helms_67,Gold_68,Voelcker_70}.  
A representative
approach~\cite{Voelcker_70} replaces the recursive computation with a
series of truncated FIR segments, reducing the per-sample complexity
from $\mathcal{O}(N)$ (linear convolution) to
$\mathcal{O}(\log_{2}N)$ (circular convolution).  
Although appealing in complexity, these FFT-based schemes approximate the
original IIR transfer function, and, as noted
in~\cite{Read_71,Meek_72}, the stability of the resulting filter is
difficult to guarantee.

A fundamentally different strategy---referred to herein as \emph{block
filtering}---was introduced by Burrus~\cite{Burrus_71,Burrus_72}.
Rather than approximating the impulse response, Burrus reformulated
the recursive filter in matrix form so that a block of output samples
can be computed jointly, converting the sample-level feedback into a
block-level recurrence.  Subsequent work exploited concurrency
features of VLSI architectures, such as pipelining, to further
accelerate block-based IIR
computation~\cite{Loomis_84,Lu_85}. 
These developments established a key
principle: the parallelism available in the target hardware should
guide the algorithmic design.

Following this principle, Sung and Mitra~\cite{Sung_86,Sung_92} were
among the first to investigate block filtering on a hypothetical
single-instruction-multiple-data (SIMD) machine.  Specialized SIMD
vector DSP architectures were subsequently
proposed~\cite{Schaffer_03,Robelly_04,Hosemann_04,Horst_05}, but
their reliance on custom hardware limited portability.  
Later efforts
targeted commodity CPUs with short-vector SIMD
instructions~\cite{Kutil_08,Ahn_09,Lee_10,Zhai_24}, expanding from
single-core to multi-core environments.  A common thread in this body
of work is that the algorithms are developed for the direct-form
realization of a general $M$-th order recursive filter.

An alternative line of research sidesteps the cascade dependency
entirely by adopting a \emph{parallel} realization, in which the
transfer function is decomposed as a sum of second-order sections
via partial fraction expansion.  The independent sections are
straightforward to distribute across processing units, as
demonstrated by Belloch et~al.~\cite{Belloch_14} for multi-channel
audio equalization on a GPU.  In this form, however, the zeros rely
on cancellation between sections and are difficult to maintain.
In practice, the \emph{cascade} of second-order sections
(biquads) remains the dominant implementation strategy.
Because the transfer function of the cascade is the product of the 
transfer functions of the biquads,
both poles and zeros are controlled accurately;
each biquad confines the effect of
coefficient quantization to a single pair of poles and zeros, 
avoiding the severe
rounding-error amplification that plagues direct-form realizations of
high-order filters~\cite[Section 6.8]{Oppenheim_10}.  
Despite these practical
advantages, the prior literature on parallel IIR algorithms has
focused almost exclusively on the direct form, leaving the additional
complexity introduced by cascading biquads
largely unaddressed.

This paper fills that gap.  We propose a block-matrix reformulation of
the cascaded second-order IIR filter that exposes parallelism at
multiple levels.  The main contributions are as follows.

\begin{enumerate}
\item \textbf{Block-matrix problem formulation:}  We cast 
  second-order recursive filtering as a banded block-Toeplitz linear
  system.  A stride-$N$ permutation then maps a group of $NL$ samples
  into a block-tridiagonal structure whose entries are scalar multiples
  of the sparse $L\times L$ identity and shift matrices.  This formulation
  provides a systematic algebraic framework in which 
  new classes of filtering algorithms
  arise naturally.
  We introduce two such algorithms that achieve,
  to the best of our knowledge,
  the highest throughput 
  reported to date on single cores of general-purpose processors.

\item \textbf{PH factorization:}  A partial LU (PH)
  factorization of the block system that preserves the sparse
  block structure throughout is introduced.
  It avoids the dense matrices that appear
  in the standard LU decomposition.  
  Within the block-matrix
  framework, it extends and generalizes the particular/homogeneous decomposition
  of~\cite{Ahn_09}.

\item \textbf{Cyclic reduction:}  We adapt cyclic reduction
 -- a classical technique for solving tridiagonal linear
  systems~\cite{Walter_97} -- to the second-order block-tridiagonal
  structure arising from the IIR filter difference equation.  
  The adaptation reduces the sequential
  dependency depth from $\mathcal{O}(N)$ in the PH factorization to
  $\mathcal{O}(\log_{2}N)$.
  Our experimental results confirm that it
  outperforms all algorithms considered herein for large sample blocks. 

\item \textbf{Cascade cost amortization:}  Existing parallel IIR
  algorithms address only a single direct-form $M$-th order filter and do not
  consider their cascaded second-order realization. 
  We demonstrate that the permutation and de-permutation required to 
  enable the block-matrix formulation is performed only once
  as permuted block matrices are passed between second-order sections in the cascade.
  As such, the overhead for permutation is amortized over the $K$ biquads in the cascade.

\item \textbf{Detailed complexity analysis and experimental
  validation:} 
  In addition to exact block-FMA and shuffle operation counts for
every algorithmic stage, we validate the theoretical predictions
against cycle-accurate measurements on three 
micro-architectures,
reporting clock cycles per sample, instructions per cycle,
and resource-stall statistics.
This analysis accounts for the impact of pipelining and multiple execution units on modern micro-processors.
We also verify that the numerical
accuracy of all algorithms is 
consistent with the predictions
of a classical round-off model.
\end{enumerate}

This paper develops the
algorithmic theory and evaluates the proposed methods on a single CPU
core using SIMD vector instructions.  
A forthcoming companion paper~\cite{zhai2026parallelcascadedrecursivefiltering} 
addresses the
implementation of the proposed algorithms on modern parallel computing
architectures, including multi-core CPUs and GPUs, targeting both
high-throughput batched processing and low-latency real-time
scenarios.

This paper is organized as follows.  Three short sections
establish the problem context, the sequential dependency
problem, and pertinent results from block filtering.
Sections~\ref{sec:multi_block_filtering} and~\ref{sec:recursive}
contain the principal contributions: the reformulation of a
signal block group as a block-tridiagonal system, and two
computationally efficient, parallel algorithms for its
solution---PH factorization and cyclic reduction.
Section~\ref{sec:permutation} addresses the permutation of
samples before and after filtering and the realization of higher-order filters.
Section~\ref{sec:results} verifies the numerical accuracy of
the proposed algorithms and demonstrates significant throughput
improvements over standard, sequential implementations.
Section~\ref{sec:conclusion} provides a summary and conclusions.
\section{The Cascaded Recursive Filter}
\label{sec:cascade}

\noindent To maintain accurate pole and zero locations,
an order $M$ recursive filter is usually realized as a cascade of
$K=\lceil M/2 \rceil$ second-order sections (biquads)
\begin{equation}
    \label{eq:general_cascaded_equation}
    y[n] \;=\; \big( F\ssub{K} \circ F\ssub{K-1}
                      \circ \cdots \circ F\ssub{1} \big)(x)[n].
\end{equation}
Each biquad takes input $x\ssub{k}[n]$ and produces
output $y\ssub{k}[n] = F\ssub{k}(x\ssub{k})[n]$ according to
the difference equation
\begin{align}
\label{eq:complex_second_order}
y\ssub{k}[n] 
&\;=\; b\ssub{0,k}\,x\ssub{k}[n] + b\ssub{1,k}\,x\ssub{k}[n-1] + b\ssub{2,k}\,x\ssub{k}[n-2] \nonumber\\
&\quad - a\ssub{1,k}\,y\ssub{k}[n-1] - a\ssub{2,k}\,y\ssub{k}[n-2]
\end{align}
with the interconnections $x\ssub{1}[n]=x[n]$, $x\ssub{k}[n]=y\ssub{k-1}[n]$ for $k=2,\ldots,K$ and $y[n]=y\ssub{K}[n]$. 
When focusing on a single biquad, we will drop the subscript $k$ from the notation.
We will also assume that
the common factor $\prod_{k=1}^K b\ssub{0,k}$ is split off so that $b\ssub{0,k}=1$ 
and the coefficients $b\ssub{1,k}$ and $b\ssub{2,k}$ are scaled by $1/b\ssub{0,k}$;
our algorithms are easily adapted if this is not desired.

\section{The Sequential Dependency Problem}
\label{sec:dependency}

\noindent Computing the scalar recurrence \eqref{eq:complex_second_order} requires 4 fused multiply-adds (FMAs) per output sample. 
Moreover, the recurrence cannot be easily parallelized, as each output depends on its two immediate predecessors, forming a sequential dependency chain across all samples.

Sequential dependency is a recurring theme in this paper.
It is of concern in high-performance implementations because such dependencies 
lead to inefficient use of the execution pipelines in modern processors.
The exact impact of sequential dependencies on performance is very difficult to model and predict.
In this paper, we rely on the concept of \emph{sequential dependency depth} (or just \emph{sequential depth}) 
to flag potential performance problems in recursive filtering algorithms.
Sequential depth describes the length of a sequence of computations that must be performed in order.
For example, to compute $N$ outputs $y[n]$ of a recursive filter via the filter's difference equation the dependency chain has length $N$.
Conversely, for a non-recursive filter,
samples $y[n]$ can be computed independently and
the summation $y[n] = \sum_{k=0}^{M} x[n-k]$ can be organized as a tree of pairwise addition resulting in a dependency depth of $\lceil \log_2(M) \rceil$. 
We will track sequential depth throughout this paper and show that it is a predictor for problems with effective pipelining in Section~\ref{sec:results}.

To expose opportunities for parallel execution, the second order recursive filter in \eqref{eq:complex_second_order} can be reformulated as a system of linear equations,
which has the matrix form
\begin{equation}
\label{eq:recursive_equation_block_form}
\Ab\,\yb 
= \Bb\,\xb
+ \Bbm\,\xb\ssub{-1} 
- \Abm\,\yb\ssub{-1},
\end{equation}
where $\Ab$ and $\Bb$
are lower-triangular banded Toeplitz
matrices with first column $[1, a\ssub{1}, a\ssub{2}, 0, \ldots]^\top$ and 
row $[1, b\ssub{1}, b\ssub{2}, 0, \ldots]^\top$, respectively,
and   
\begin{equation}
\label{eq:filtering_ABm}
\Abm = 
\begin{bmatrix}
a\ssub{2} & a\ssub{1} \\
& a\ssub{2} \\
& \\
\end{bmatrix}, \quad
\Bbm = 
\begin{bmatrix}
b\ssub{2} & b\ssub{1} \\
& b\ssub{2} \\
& \\
\end{bmatrix},
\end{equation}
with $\xb\ssub{-1} = [x[-2],\, x[-1]]^\top$ and 
$\yb\ssub{-1} = [y[-2],\, y[-1]]^\top$. 
Then \eqref{eq:complex_second_order}
is just the sequential solution of \eqref{eq:recursive_equation_block_form} by forward substitution. 
However, the matrix formulation enables alternative solution methods
that facilitate parallel processing.
Such methods are the focus of this paper.

\section{Block Filtering}
\label{sec:block_filtering}

\noindent A classical approach to exposing the parallelism in~\eqref{eq:recursive_equation_block_form} is to partition the input and output sequences into contiguous, non-overlapping blocks \cite{Burrus_71}.
Let $L$ denote the \emph{block size} and assume 
that $L$ is at least the filter order, i.e., $L \geq 2$. Applying partitioning
to~\eqref{eq:recursive_equation_block_form} with block index~$n$
yields
\begin{equation}
    \label{eq:block_filtering_equation_An}
    \Ab\,\yb\ssub{n}
      = \Bb\,\xb\ssub{n}
      + \Bbm\,\xbm\ssub{n-1}
      - \Abm\,\ybm\ssub{n-1},
\end{equation}
where
$\xb\ssub{n}
  = [x[nL],\, x[nL+1],\, \ldots,\, x[(n+1)L-1]]^\top$
denotes the $n$-th input block (with $\yb\ssub{n}$ defined
analogously).
$\xbm\ssub{n-1}
  = [x[nL-2],\, x[nL-1]]^\top$ and
$\ybm\ssub{n-1}
  = [y[nL-2],\, y[nL-1]]^\top$
contain the two terminal samples of the preceding block.
$\Ab$, $\Bb$ are  $L \times L$  sub-matrices
of the matrices in~\eqref{eq:recursive_equation_block_form}, and $\Abm$, $\Bbm$ are the
corresponding $L \times 2$ sub-matrices
of~\eqref{eq:filtering_ABm}.

Since $\Ab$ is lower-triangular and thus always
invertible, \eqref{eq:block_filtering_equation_An} can be  solved directly by
\begin{equation}
    \label{eq:block_filtering_equation}
    \yb\ssub{n} = \Hb\,\xb\ssub{n}
        + \Hbb\,\xbm\ssub{n-1}
        - \Hba\,\ybm\ssub{n-1},
\end{equation}
where 
\begin{equation}
\label{eq:block_filtering_matrices}
    \Hb = \Ab^{-1}\,\Bb, \quad
    \Hbb = \Ab^{-1}\,\Bbm, \quad
    \Hba = \Ab^{-1}\,\Abm.
\end{equation}
The matrix $\Hb$ is a dense lower-triangular Toeplitz matrix, filled with the first $L$ samples of the impulse response \cite{Burrus_71}. 
Equation \eqref{eq:block_filtering_equation} extends to a block recurrence, 
where each length~$L$ block depends on the two terminal output samples from the preceding block.

In this work, we treat blocks of $L$ samples as the
fundamental unit of computation, with~$L$ chosen to match the native
parallelism of the hardware, e.g., the SIMD register width on a CPU
or the warp width on a GPU. 

We define a \emph{block FMA} as a fused multiply-add applied
element-wise to $L$-sample blocks, so that one block FMA
performs $L$ scalar FMAs in parallel.
Computing~\eqref{eq:block_filtering_equation} for a single
block is a cumulative sum of $L$ column contributions from
the dense matrix~$\Hb$.
This requires $L$ block FMAs plus
$4$~block FMAs for the boundary corrections $\Hbb$ and
$\Hba$ for a total of $L + 4$ block FMAs per block, or
equivalently, $1 + 4/L$ block FMAs per sample,
compared to
$4$~FMAs per sample in the scalar
recurrence~\eqref{eq:complex_second_order}.
\section{Multi-Block Filtering}
\label{sec:multi_block_filtering}

\noindent Multi-block filtering extends the block recurrence \eqref{eq:block_filtering_equation_An} by
considering larger groups of~$NL$ samples, where~$N$ denotes the
\emph{number of blocks}, and~$L$ is still the block size. 
A group of~$NL$
samples is referred to as a \emph{signal block group}~\cite{Ahn_09}.
By processing $N$ blocks jointly, 
the method creates opportunities to restructure the inter-block recurrence and exploit parallelism across blocks beyond the parallelism available within each block in block filtering.

To enable parallel processing of $N$ blocks within a signal block group,
the~$NL$ samples are first permuted with a stride of~$N$ (see Section~\ref{sec:permutation}).
Then, the multi-block filtering equation for a
\emph{single} signal block group yields
\begin{equation}
\label{eq:recursive_equation_multi_block_form}
\Ag\,\yg
= \Bg\,\xg 
+ \Bgm\,\xb\ssub{-1} 
- \Agm\,\yb\ssub{-1},
\end{equation}
where
$\xg = [\xbp\ssub{0}^\top,\, \xbp\ssub{1}^\top,\, \ldots,\,
\xbp\ssub{N-1}^\top]^\top$ and $\xbp\ssub{m}$ denotes the $m$-th
strided block in a signal block group, defined as
$\xbp\ssub{m} = [x[m],\, x[m{+}N],\, \ldots,\,
x[m{+}(L{-}1)N]]^\top$,
$\yg$ is defined analogously,
and $\xb\ssub{-1} = [x[-2],\, x[-1]]^\top$,
$\yb\ssub{-1} = [y[-2],\, y[-1]]^\top$ 
denote the initial state provided by the prior signal block group. 
After permutation,
the system matrix for the signal block group becomes
\begin{equation}
\label{eq:multi_block_A}
\Ag = 
\begin{bmatrix}
\bm{I}          &                    &                    &  a\ssub{2}\SR & a\ssub{1}\SR \\
a\ssub{1}\bm{I} & \bm{I}         &                    &                     & a\ssub{2}\SR \\
a\ssub{2}\bm{I} & a\ssub{1}\bm{I}& \bm{I}         &                     &                     \\
                     & \ddots             & \ddots             & \ddots              &                     \\
                     &                    & a\ssub{2}\bm{I}& a\ssub{1}\bm{I} & \bm{I}
\end{bmatrix},
\end{equation}
where $\Ag \in \mathbb{R}^{NL \times NL}$ has a block Toeplitz structure
with $L \times L$ entries: each sub-matrix is a scalar multiple
of the identity matrix~$\bm{I}$ or the lower shift matrix~$\SR$,
which only has ones on the first sub-diagonal. 
The diagonal $\bm{I}$~blocks represent the intra-block identity, the
sub-diagonal $a\ssub{1}\bm{I}$ and $a\ssub{2}\bm{I}$ blocks capture
the causal dependencies between two consecutive blocks, and the
upper-corner blocks $a\ssub{1}\SR$ and $a\ssub{2}\SR$ couple
the final two samples from prior blocks to
the first two blocks. 
The input matrix $\Bg$ has the same structure as $\Ag$ with
$a\ssub{1}, a\ssub{2}$ replaced by $b\ssub{1}, b\ssub{2}$.
The corresponding terminal matrices
$\Agm, \Bgm \in \mathbb{R}^{NL \times 2}$ are
\begin{equation}
\label{eq:multi_block_Am}
\Agm = 
\begin{bmatrix}
a\ssub{2}\Se & a\ssub{1}\Se \\
                              & a\ssub{2}\Se \\
                              &                              \\
\end{bmatrix}, \quad
\Bgm = 
\begin{bmatrix}
b\ssub{2}\Se & b\ssub{1}\Se \\
                              & b\ssub{2}\Se \\
                              &                              \\
\end{bmatrix},
\end{equation}
where $\Se = [1,\, 0,\, \ldots,\, 0]^\top \in \mathbb{R}^{L}$.

Recall that the block recurrence extends to $NL$ samples in multi-block filtering. 
Consequently, we focus on a single signal block group and
present~\eqref{eq:recursive_equation_multi_block_form} with
local indices only, i.e., without reference to the absolute sample indices.

The computation of \eqref{eq:recursive_equation_multi_block_form}  separates naturally into two stages. 
The non-recursive stage computes the right-hand-side vector
\begin{equation}
\label{eq:feedforward}
\vg = \Bg\,\xg 
+ \Bgm\,\xb\ssub{-1}
- \Agm\,\yb\ssub{-1},
\end{equation}
where 
$\vg = [\vbp\ssub{0}^{\top},\, \vbp\ssub{1}^{\top},\, \ldots,\, \vbp\ssub{N-1}^{\top}]^\top$ 
depends only on the input blocks and the  initial state. 
The recursive stage then solves
\begin{equation}
\label{eq:recursive}
\Ag\,\yg = \vg.
\end{equation}
We treat the two stages separately, beginning with the non-recursive part.
Solving the recursive problem is the main contribution
of this paper and is addressed in Section~\ref{sec:recursive}.

\subsection{The Non-Recursive Stage}
\label{sec:non-recursive}

Expanding \eqref{eq:feedforward} on a per-block basis yields
\begin{equation}
\label{eq:feedforward_block}
\begin{aligned}
\vbp\ssub{0} =& \xbp\ssub{0} 
{+} b\ssub{1}\SR\,\xbp\ssub{N-1} 
{+} b\ssub{2}\SR\,\xbp\ssub{N-2} {+}\\
&b\ssub{1}\Se x[-1] 
{+} b\ssub{2}\Se x[-2]
{-} a\ssub{1}\Se y[-1] 
{-} a\ssub{2}\Se y[-2], \\
\vbp\ssub{1} =& \xbp\ssub{1} 
+ b\ssub{1}\xbp\ssub{0} 
+ b\ssub{2}\SR\,\xbp\ssub{N-1} {+}\\
& b\ssub{2}\Se x[-1]
- a\ssub{2}\Se y[-1], \\
\vbp\ssub{m} =& \xbp\ssub{m} 
+ b\ssub{1}\xbp\ssub{m-1} 
+ b\ssub{2}\xbp\ssub{m-2} 
\text{, $m = 2, \ldots, N-1$}.
\end{aligned}
\end{equation}
For $m \geq 2$, each block depends only on its two immediate
predecessors. 
The first two blocks ($m = 0, 1$) are additionally
coupled to the initial state through the shift matrix $\SR$,
the unit vector $\Se$, and the initial output samples $y[-1],\, y[-2]$. 
Since $\SR\,\xbp\ssub{N-1} = [0,\,x[N-1],\,x[2N-1],\,\ldots]^\top$ shifts a block down by one position, and
$\Se x[-1] = [x[-1],\,0,\,0,\,\ldots]^\top$ selects only the first entry, 
we can define extended input blocks $\xbp\ssub{-1} = [x[-1],\, x[N-1],\,
\ldots,\, x[(L-1)N-1]]^\top$ and
$\xbp\ssub{-2} = [x[-2],\, x[N-2],\,
\ldots,\, x[(L-1)N-2]]^\top$ that absorb the input boundary terms.
Then \eqref{eq:feedforward_block} reduces to
\begin{equation}
\label{eq:feedforward_block_uniform}
\begin{aligned}
\vbp\ssub{m} =& \xbp\ssub{m} 
+ b\ssub{1}\xbp\ssub{m-1} 
+ b\ssub{2}\xbp\ssub{m-2} 
\text{, $m = 0, \ldots, N{-}1$,} \\
\vbp\ssub{0} \leftarrow &
\vbp\ssub{0}
- a\ssub{1}\Se\, y[-1] 
- a\ssub{2}\Se\, y[-2], \\
\vbp\ssub{1} \leftarrow &
\vbp\ssub{1}
- a\ssub{2}\Se\, y[-1].
\end{aligned}
\end{equation}
This expression consists of a block-wise FIR equation that can be computed in parallel by multiple execution units
across all~$N$ blocks, followed by scalar corrections to the first element of the first two blocks. 
The cost of the non-recursive stage is $2N$ block FMAs,
$2$~shuffles for constructing the extended input blocks, and $3$~additional scalar FMAs to absorb the initial output state into the first two blocks.
\section{Parallel Algorithms for the Recursive Stage}
\label{sec:recursive}

\noindent With $\vg$ computed, the problem reduces to solving the recursive
stage~\eqref{eq:recursive}. 
Direct solution via the dense matrix $\Ag^{-1}$ is as expensive as block filtering.
Instead, we seek more efficient ways to solve~\eqref{eq:recursive},
including factorizations of $\Ag$ and cyclic reduction.

\subsection{LU Factorization}
\label{sec:LU}

An obvious approach to solving~~\eqref{eq:recursive}  
factors $\Ag$ into a lower triangular matrix $\bm{L}$ 
and an upper triangular matrix $\bm{U}$, that is,
\begin{equation}
\label{eq:LU_factorization}
\Ag=\bm{L}\bm{U}=\left[
\begin{array}{c:c}
\bm{L}\ssub{11} & \bm{0} \\
\hdashline
\bm{L}\ssub{21} & \bm{L}\ssub{22} \\
\end{array}
\right]
\left[
\begin{array}{c:c}
\bm{U}\ssub{11} & \bm{U}\ssub{12} \\
\hdashline
\bm{0} & \bm{U}\ssub{22} \\ 
\end{array}
\right],
\end{equation}
where $\bm{U}\ssub{11} = \bm{I}$ and 
$\bm{L}\ssub{11} \in \mathbb{R}^{(N-2)L \times (N-2)L}$ is lower triangular, block Toeplitz with first column equal to 
$[\bm{I}, a\ssub{1}\bm{I}, a\ssub{2}\bm{I}, \bm{0},\ldots,\bm{0}]^\top$.
The lower left matrix $\bm{L}\ssub{21} \in \mathbb{R}^{2L \times (N-2)L}$
extends this pattern to $N$ blocks,
\begin{equation}
\bm{L}\ssub{21} =
\begin{bmatrix}
 \bm{0} & \ldots & \bm{0} & a\ssub{2}\bm{I} & a\ssub{1}\bm{I} \\
 \bm{0} & \ldots & \bm{0} & \bm{0} & a\ssub{2}\bm{I}
\end{bmatrix},
\end{equation}
and $\bm{U}\ssub{12} \in \mathbb{R}^{(N-2)L \times 2L}$
captures the coupling introduced by 
the shift matrix,
\begin{equation}
\label{eq:U_12}
\bm{U}\ssub{12} = 
\begin{bmatrix}
u\ssub{2,0}\SR & u\ssub{1,0}\SR \\
u\ssub{2,1}\SR & u\ssub{1,1}\SR \\
\vdots & \vdots \\
u\ssub{2,N-3}\SR & u\ssub{1,N-3}\SR
\end{bmatrix}.
\end{equation}
Here $u\ssub{2,0} = a\ssub{2}$, $u\ssub{1,0} = a\ssub{1}$, 
$u\ssub{2,1} = -a\ssub{1}a\ssub{2}$, 
$u\ssub{1,1} = a\ssub{2} - a\ssub{1}^2$, and the remaining 
entries follow the recurrence 
$u\ssub{l,m} = -a\ssub{2}\,u\ssub{l,m-2} - a\ssub{1}\,u\ssub{l,m-1}$ 
for $m \geq 2$.


The key structural difference appears
in the lower right matrices 
\begin{equation}
\label{eq:L_22_U_22_inside_matrices}
\bm{L}\ssub{22}=
\begin{bmatrix}
\bm{I}  & \bm{0} \\
\bm{K}  & \bm{I}
\end{bmatrix} \text{ and }
\bm{U}\ssub{22}=
\begin{bmatrix}
\bm{I}+\bm{E} & \bm{G} \\
\bm{0} & \bm{I}+\bm{J}
\end{bmatrix}
\end{equation}
with
\begin{equation}
\begin{aligned}
\bm{E} &= u\ssub{2,N-2}\,\SR, \\
\bm{G} &= u\ssub{1,N-2}\,\SR, \\
\bm{K} &= (a\ssub{1}\bm{I} 
  - a\ssub{2}\,u\ssub{2,N-3}\,\SR)
  (\bm{I} + \bm{E})^{-1}, \\
\bm{J} &= -a\ssub{2}\,u\ssub{1,N-3}\,\SR 
  - \bm{K}\bm{G}.
\end{aligned}
\end{equation}
Note that $(\bm{I} + \bm{E})^{-1}$ is a dense lower 
triangular Toeplitz matrix and 
$(a\ssub{1}\bm{I} - a\ssub{2}\,u\ssub{2,N-3}\,\SR)$ 
is banded.
Hence, their product $\bm{K}$ is also a dense lower triangular 
Toeplitz matrix. Moreover, $\bm{J}$ is also dense lower 
triangular Toeplitz because it depends on $\bm{K}$.

The factorization $\Ag = \bm{L}\bm{U}$ transforms the
recursive stage~\eqref{eq:recursive} into a forward substitution
$\bm{L}\,\wg = \vg$, followed by a back substitution
$\bm{U}\,\yg = \wg$. 
Each requires $N-2$ substitutions involving $\bm{I}$
and $\SR$ that are solved efficiently with block superpositions and block shuffles.
However, solving the blocks involving $\bm{L}\ssub{22}$ and $\bm{U}\ssub{22}$ includes dense
matrices $\bm{K}$ and $\bm{J}$.
Their solution is inefficient, as each incurs $L$ block FMAs per
block. 
A detailed complexity analysis is deferred until we can compare it to the more efficient alternative discussed next.

\subsection{PH Factorization}
\label{sec:PH}

The LU factorization of $\Ag$ is only partially efficient due to the presence of dense matrices $\bm{K}$ in $\bm{L\ssub{22}}$ and $\bm{J}$ in $\bm{U\ssub{22}}$. 
These dense structures arise because, in the standard LU decomposition, the matrix $\bm{U\ssub{22}}$
is strictly block upper triangular to allow backward substitution. 
As a result, off-diagonal terms $a\ssub{1}\bm{R}$ and $a\ssub{2}\bm{R}$ in $\Ag$ 
produce the dense lower triangular matrix $\bm{K}$ and, in turn, the dense structure $\bm{J}$.

A more favorable factorization of $\Ag$ arises when the upper-triangular constraint is relaxed. 
Define the partial LU factorization  as
\begin{equation}
\label{eq:PH_decomposition}
\Ag =\bm{P}\bm{H}=\bm{P}
\left[
\begin{array}{c: c}
\bm{H\ssub{11}} & \bm{H\ssub{12}} \\
\hdashline
\bm{0} & \bm{H\ssub{22}} \\
\end{array}
\right],
\end{equation}
where $\bm{P} \in \mathbb{R}^{NL \times NL}$ extends the block Toeplitz form of $\bm{L\ssub{11}}$ to $NL \times NL$, 
so that forward substitution is trivial and efficient.

The matrices $\bm{H\ssub{11}}$ and $\bm{H\ssub{12}}$ are equal to $\bm{U\ssub{11}}$ and $\bm{U\ssub{12}}$, respectively.
Matrix $\bm{H\ssub{22}}$, in contrast to $\bm{U\ssub{22}}$ in \eqref{eq:L_22_U_22_inside_matrices}, is no longer constrained to be upper triangular,
\begin{equation}
\label{eq:H_22}
\bm{H\ssub{22}}=
\begin{bmatrix}
\bm{I}+e\SR & g\SR \\
f\SR & \bm{I}+d\SR
\end{bmatrix},
\end{equation}
where the scalars $e = u\ssub{2,N-2}$, $g = u\ssub{1,N-2}$,
$f = u\ssub{2,N-1}$, and $d = u\ssub{1,N-1}$ extend the
pattern of the coefficients in~\eqref{eq:U_12}. 
Crucially, the entire factorization $\Ag = \bm{P}\bm{H}$ preserves the sparse block structure 
consisting only of
scaled identity and shift matrices
$\bm{I}$ and $\SR$.

\subsubsection{Particular Solution and Homogeneous Correction}

Substituting $\Ag = \bm{P}\bm{H}$ into~\eqref{eq:recursive},
the recursive stage is computed in two steps:
\begin{equation}
\label{eq:particular_and_homogeneous}
\bm{P}\,\wg = \vg
\text{ followed by }
\bm{H}\,\yg = \wg.
\end{equation}
The forward substitution in~\eqref{eq:particular_and_homogeneous}
yields $\wg$ through a second-order block recurrence.
For the back-substitution in~\eqref{eq:particular_and_homogeneous},
since $\bm{H}\ssub{11} = \bm{I}$ and
$\bm{H}\ssub{12} = \bm{U}\ssub{12}$
in~\eqref{eq:U_12}, 
the blocks $\ybp\ssub{m}$ for $m=0, \ldots, N-3$
depend only on the two terminal blocks and are given by
\begin{equation}
\label{eq:PH_forward}
\ybp\ssub{m}
= \wbp\ssub{m}
- u\ssub{2,m}\,\SR\,\ybp\ssub{N-2}
- u\ssub{1,m}\,\SR\,\ybp\ssub{N-1}.
\end{equation}

The partial LU factorization thus leads to a
\textit{particular--homogeneous decomposition}: the forward
substitution in~\eqref{eq:particular_and_homogeneous} computes the
\textit{particular solution} under zero initial conditions,
and~\eqref{eq:PH_forward} applies the \textit{homogeneous
correction} using the (yet to be determined) terminal blocks
that include the initial conditions. 
A similar decomposition was
introduced by Ahn~\cite{Ahn_09} for first-order recursive
equations.
The present work extends it to second-order sections
and introduces the block matrix framework.
We will show below that the factorization also allows us to find the terminal blocks efficiently.
The name
\textit{PH factorization} is used to distinguish it from the standard LU
factorization and to emphasize this connection.

To initiate back-substitution~\eqref{eq:PH_forward}, 
it remains to determine the two terminal blocks
$\ybp\ssub{N-2}$ and $\ybp\ssub{N-1}$, which are governed
by the $2L \times 2L$ block system $\bm{H}\ssub{22}$
in~\eqref{eq:H_22}. 
Direct solution
requires inverting $(\bm{I} + e\SR)$ and $(\bm{I} + d\SR)$;
both of these inverses are dense lower triangular Toeplitz---the same
costly structure that the PH factorization was supposed to avoid.
Instead, a more efficient method is developed below.

\subsubsection{Solving the Terminal Blocks by Recursive Doubling}

To compute the two terminal blocks $\ybp\ssub{N-2}$ and
$\ybp\ssub{N-1}$ governed by $\bm{H}\ssub{22}$
in~\eqref{eq:H_22}, we first expand the $2L \times 2L$ system into
its sample-level representation,
\begin{equation*}
\label{eq:recursive_doubling_before_permute}
\scalebox{0.85}{$
\left[
\begin{array}{cccc:cccc}
 1 &   &   &   &   &   &   &   \\
 e &  1 &   &   & g  &   &   &   \\
  & \ddots  & \ddots  &   &   & \ddots  &   &   \\
  &   & e  & 1  &   &   & g  &   \\
\hdashline
  &   &   &   & 1  &   &   &   \\
 f &   &   &   & d  & 1  &   &   \\
  & \ddots  &   &   &   & \ddots  & \ddots  &   \\
  &   & f  &   &   &   & d  & 1 \\
\end{array}
\right]
\begin{bmatrix}
 y[N-2] \\ y[2N-2] \\ \vdots \\ y[LN-2] \\
\hdashline
 y[N-1] \\ y[2N-1] \\ \vdots \\ y[LN-1] \\
\end{bmatrix} 
=
\begin{bmatrix}
 w[N-2] \\ w[2N-2] \\ \vdots \\ w[LN-2] \\
\hdashline
 w[N-1] \\ w[2N-1] \\ \vdots \\ w[LN-1] \\
\end{bmatrix}.
$}
\end{equation*}
Permuting equations by grouping consecutive samples yields
\begin{equation}
\label{eq:recursive_doubling_after_permute}
\begin{bmatrix}
\bm{I}  &              &             &    \\
\bm{C}  & \bm{I}   &             &    \\
            & \ddots   & \ddots  &    \\
            &              & \bm{C}  & \bm{I}
\end{bmatrix}
\begin{bmatrix}
\ye[0] \\ 
\ye[1] \\ 
\vdots \\ 
\ye[L-1] 
\end{bmatrix} 
=
\begin{bmatrix}
\we[0] \\ 
\we[1] \\ 
\vdots \\ 
\we[L-1] 
\end{bmatrix} 
\end{equation}
where
$\ye[l] = [y[(l+1)N-2],\, y[(l+1)N-1]]^\top$
for $l = 0, 1, \ldots, L-1$,
$\we[l]$ is defined analogously, and 
\begin{equation}
\label{eq:C}
\bm{C}=
\begin{bmatrix}
 e & g \\  
 f & d 
\end{bmatrix}.
\end{equation}
Note that each row of~\eqref{eq:recursive_doubling_after_permute} forms
a $2 \times 2$ first-order matrix difference equation, 
\begin{equation}
\label{eq:recursive_doubling_equation}
\ye[l]
= \we[l] - \bm{C}\,\ye[l-1],
\;\text{for}\; l = 1, \ldots, L-1,
\end{equation}
with $\ye[0] = \we[0]$.

\begin{figure}[t]
\centering
\scalebox{0.8}{$
\begin{tikzpicture}[scale=0.71]
    
    \node[circle, fill=black, inner sep=1.5pt] (ye7) at (0, 6) {};
    \node[above=2pt] at (ye7) {$\ye[7]$};
    
    \node[circle, fill=black, inner sep=1.5pt] (ye6) at (2, 6) {};
    \node[above=2pt] at (ye6) {$\ye[6]$};
    
    \node[circle, fill=black, inner sep=1.5pt] (ye5) at (4, 6) {};
    \node[above=2pt] at (ye5) {$\ye[5]$};
    
    \node[circle, fill=black, inner sep=1.5pt] (ye4) at (6, 6) {};
    \node[above=2pt] at (ye4) {$\ye[4]$};
    
    \node[circle, fill=black, inner sep=1.5pt] (l2n1) at (0, 4) {};
    \node[circle, fill=black, inner sep=1.5pt] (l2n2) at (2, 4) {};
    
    \node[circle, fill=black, inner sep=1.5pt] (ye3) at (8, 4) {};
    \node[above=2pt, xshift=9pt] at (ye3) {$\ye[3]$};
    
    \node[circle, fill=black, inner sep=1.5pt] (ye2) at (10, 4) {};
    \node[above=2pt] at (ye2) {$\ye[2]$};
    
    \node[circle, fill=black, inner sep=1.5pt] (l3n1) at (0, 2) {};
    \node[circle, fill=black, inner sep=1.5pt] (l3n3) at (4, 2) {};
    \node[circle, fill=black, inner sep=1.5pt] (l3n5) at (8, 2) {};
    
    \node[circle, fill=black, inner sep=1.5pt] (ye1) at (12, 2) {};
    \node[right=2pt,xshift=5pt] at (ye1) {$\ye[1]$};
    
    \node[circle, fill=black, inner sep=1.5pt] (we7) at (0, 0) {};
    \node[below=2pt] at (we7) {$\we[7]$};
    
    \node[circle, fill=black, inner sep=1.5pt] (we6) at (2, 0) {};
    \node[below=2pt] at (we6) {$\we[6]$};
    
    \node[circle, fill=black, inner sep=1.5pt] (we5) at (4, 0) {};
    \node[below=2pt] at (we5) {$\we[5]$};
    
    \node[circle, fill=black, inner sep=1.5pt] (we4) at (6, 0) {};
    \node[below=2pt] at (we4) {$\we[4]$};
    
    \node[circle, fill=black, inner sep=1.5pt] (we3) at (8, 0) {};
    \node[below=2pt] at (we3) {$\we[3]$};
    
    \node[circle, fill=black, inner sep=1.5pt] (we2) at (10, 0) {};
    \node[below=2pt] at (we2) {$\we[2]$};
    
    \node[circle, fill=black, inner sep=1.5pt] (we1) at (12, 0) {};
    \node[below=2pt] at (we1) {$\we[1]$};
    
    \node[circle, fill=black, inner sep=1.5pt] (ye0) at (14, 0) {};
    \node[below=2pt] at (ye0) {$\ye[0]$};





    
    
    
    \draw[->, thick] (we7) -- (l3n1);
    \draw[->, thick] (we6) -- (l2n2);
    \draw[->, thick] (we5) -- (l3n3);
    \draw[->, thick] (we3) -- (l3n5);
    \draw[->, thick] (we1) -- (ye1);
    
    \draw[->, thick] (l3n1) -- (l2n1);
    \draw[->, thick] (l3n5) -- (ye3);
    \draw[->, thick] (we2) -- (ye2);
    
    \draw[->, thick] (l2n1) -- (ye7);
    \draw[->, thick] (l2n2) -- (ye6);
    \draw[->, thick] (l3n3) -- (ye5);
    \draw[->, thick] (we4) -- (ye4);
    
    \draw[->, thick] (we6) -- (l3n1) node[midway, above , font=\small] {$-\bm{C}$};
    
    
    \draw[->, thick] (we4) -- (l3n3) node[midway, above , font=\small] {$-\bm{C}$};
    
    
    \draw[->, thick] (we2) -- (l3n5) node[midway, above , font=\small] {$-\bm{C}$};
    
    
    \draw[->, thick] (ye0) -- (ye1) node[midway, above, font=\small] {$-\bm{C}$};
    
    
    \draw[->, thick] (l3n3) -- (l2n1) node[near end, above , font=\small] {$\bm{C}^2$};
    
    \draw[->, thick] (l3n3) -- (l2n2) node[midway, above , font=\small] {$-\bm{C}$};
    
    
    \draw[->, thick] (ye1) -- (ye2) node[midway, above , font=\small] {$-\bm{C}$};
    
    \draw[->, thick] (ye1) -- (ye3) node[near end, above , font=\small] {$\bm{C}^2$};
    
    \draw[->, thick] (ye3) -- (ye7) node[very near end, above , font=\small] {$\bm{C}^4$};
    
    \draw[->, thick] (ye3) -- (ye6) node[very near end, above , font=\small] {$-\bm{C}^3$};
    
    \draw[->, thick] (ye3) -- (ye5) node[near end, above , font=\small] {$\bm{C}^2$};
    
    \draw[->, thick] (ye3) -- (ye4) node[near end, above , font=\small] {$-\bm{C}$};
    

\end{tikzpicture}
$}
\caption{Sklansky
construction~\cite{Sklansky_60} for computing matrix
scan~\eqref{eq:recursive_doubling_equation} with $L = 8$.}
\label{fig:recursive_doubling_diagram}
\end{figure}
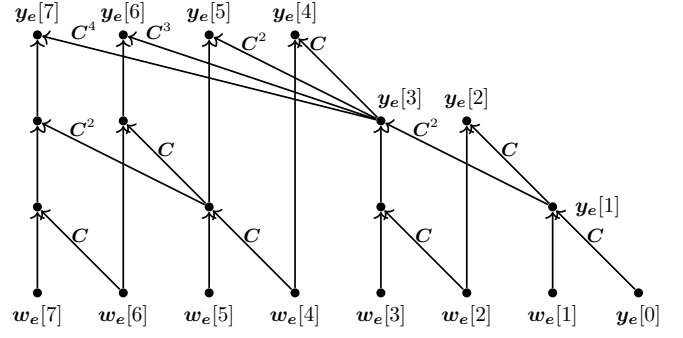

The conventional way of solving~\eqref{eq:recursive_doubling_equation}
in parallel is by recursive doubling~\cite{Kogge_73}, which
reduces the sequential depth from $\mathcal{O}(L)$
to~$\mathcal{O}(\log_2 L)$.  Merrill and
Garland~\cite{Merrill_16} compare four parallel prefix
constructions~\cite{Kogge_73,Sklansky_60,Brent_82,Yuri_08} with different depth--size trade-offs; among
them, the Sklansky construction~\cite{Sklansky_60} achieves
the minimum sequential depth with the smallest circuit size
and is adopted here; it is illustrated in
Figure~\ref{fig:recursive_doubling_diagram} for $L = 8$.

\begin{figure*}[t]
\centering
\scalebox{1.0}{%
\begin{tikzpicture}

\node[draw, align=center, font=\scriptsize, inner sep=4pt] (w0) at (0, 0) 
    {$\bm{w}\ssub{0}[0]$ \\ $\bm{w}\ssub{0}[1]$ \\ $\bm{w}\ssub{0}[2]$ \\ $\bm{w}\ssub{0}[3]$ \\ $\bm{w}\ssub{0}[4]$ \\ $\bm{w}\ssub{0}[5]$ \\ $\bm{w}\ssub{0}[6]$ \\ $\bm{w}\ssub{0}[7]$};
\node[draw, align=center, font=\scriptsize, inner sep=4pt, right=0.4em of w0] (w1)
    {$\bm{w}\ssub{1}[0]$ \\ $\bm{w}\ssub{1}[1]$ \\ $\bm{w}\ssub{1}[2]$ \\ $\bm{w}\ssub{1}[3]$ \\ $\bm{w}\ssub{1}[4]$ \\ $\bm{w}\ssub{1}[5]$ \\ $\bm{w}\ssub{1}[6]$ \\ $\bm{w}\ssub{1}[7]$};
 
\foreach \i in {1,3,5,7} {
    \draw[red, thick, rounded corners=1pt] 
        ([yshift={-0.4em-\i*0.8em}, xshift=2pt]w0.north west) 
        rectangle 
        ([yshift={-1.2em-\i*0.8em}, xshift=-2pt]w1.north east);
}
 
\node[draw, fill=blue!60!black!10, align=center, font=\scriptsize, inner sep=4pt, right=2em of w1] (k1s1)
    {$\bm{w}\ssub{0}[1]$ \\ $\bm{w}\ssub{1}[1]$ \\ $\bm{w}\ssub{0}[3]$ \\ $\bm{w}\ssub{1}[3]$ \\ $\bm{w}\ssub{0}[5]$ \\ $\bm{w}\ssub{1}[5]$ \\ $\bm{w}\ssub{0}[7]$ \\ $\bm{w}\ssub{1}[7]$};
\node[draw, fill=green!60!black!10, align=center, font=\scriptsize, inner sep=4pt, right=0.4em of k1s1] (k1v1)
    {$\bm{w}\ssub{0}[0]$ \\ $\bm{w}\ssub{1}[0]$ \\ $\bm{w}\ssub{0}[2]$ \\ $\bm{w}\ssub{1}[2]$ \\ $\bm{w}\ssub{0}[4]$ \\ $\bm{w}\ssub{1}[4]$ \\ $\bm{w}\ssub{0}[6]$ \\ $\bm{w}\ssub{1}[6]$};
 
\foreach \top in {2,6} {
    \draw[red, thick, rounded corners=1pt] 
        ([yshift={-0.4em-\top*0.8em}, xshift=2pt]k1s1.north west) 
        rectangle 
        ([yshift={-1.2em-(\top+1)*0.8em}, xshift=-2pt]k1v1.north east);
}
 
\node[draw, fill=green!60!black!10, align=center, font=\scriptsize, inner sep=4pt, 
      below=2em of k1s1] (k1p)
    {$\bm{w}\ssub{0}[0]$ \\ $\bm{w}\ssub{1}[0]$ \\ $\bm{w}\ssub{0}[2]$ \\ $\bm{w}\ssub{1}[2]$ \\ $\bm{w}\ssub{0}[4]$ \\ $\bm{w}\ssub{1}[4]$ \\ $\bm{w}\ssub{0}[6]$ \\ $\bm{w}\ssub{1}[6]$};
\node[draw, fill=green!60!black!10, align=center, font=\scriptsize, inner sep=4pt, right=0.4em of k1p] (k1q)
    {$\bm{w}\ssub{1}[0]$ \\ $\bm{w}\ssub{0}[0]$ \\ $\bm{w}\ssub{1}[2]$ \\ $\bm{w}\ssub{0}[2]$ \\ $\bm{w}\ssub{1}[4]$ \\ $\bm{w}\ssub{0}[4]$ \\ $\bm{w}\ssub{1}[6]$ \\ $\bm{w}\ssub{0}[6]$};
 
\node[draw, fill=blue!60!black!10, align=center, font=\scriptsize, inner sep=4pt, right=2em of k1v1] (k2s1)
    {$\bm{w}\ssub{0}[2]$ \\ $\bm{w}\ssub{1}[2]$ \\ $\bm{w}\ssub{0}[3]$ \\ $\bm{w}\ssub{1}[3]$ \\ $\bm{w}\ssub{0}[6]$ \\ $\bm{w}\ssub{1}[6]$ \\ $\bm{w}\ssub{0}[7]$ \\ $\bm{w}\ssub{1}[7]$};
\node[draw, fill=green!60!black!10, align=center, font=\scriptsize, inner sep=4pt, right=0.4em of k2s1] (k2v1)
    {$\bm{w}\ssub{0}[0]$ \\ $\bm{w}\ssub{1}[0]$ \\ $\bm{w}\ssub{0}[1]$ \\ $\bm{w}\ssub{1}[1]$ \\ $\bm{w}\ssub{0}[4]$ \\ $\bm{w}\ssub{1}[4]$ \\ $\bm{w}\ssub{0}[5]$ \\ $\bm{w}\ssub{1}[5]$};
 
\draw[red, thick, rounded corners=1pt] 
    ([yshift={-0.4em-4*0.8em}, xshift=2pt]k2s1.north west) 
    rectangle 
    ([yshift={-1.2em-7*0.8em}, xshift=-2pt]k2v1.north east);
 
\node[draw, fill=green!60!black!10, align=center, font=\scriptsize, inner sep=4pt, 
      below=2em of k2s1] (k2p)
    {$\bm{w}\ssub{0}[1]$ \\ $\bm{w}\ssub{1}[1]$ \\ $\bm{w}\ssub{0}[1]$ \\ $\bm{w}\ssub{1}[1]$ \\ $\bm{w}\ssub{0}[5]$ \\ $\bm{w}\ssub{1}[5]$ \\ $\bm{w}\ssub{0}[5]$ \\ $\bm{w}\ssub{1}[5]$};
\node[draw, fill=green!60!black!10, align=center, font=\scriptsize, inner sep=4pt, right=0.4em of k2p] (k2q)
    {$\bm{w}\ssub{1}[1]$ \\ $\bm{w}\ssub{0}[1]$ \\ $\bm{w}\ssub{1}[1]$ \\ $\bm{w}\ssub{0}[1]$ \\ $\bm{w}\ssub{1}[5]$ \\ $\bm{w}\ssub{0}[5]$ \\ $\bm{w}\ssub{1}[5]$ \\ $\bm{w}\ssub{0}[5]$};
 
\node[draw, fill=blue!60!black!10, align=center, font=\scriptsize, inner sep=4pt, right=2em of k2v1] (bk1s1)
    {$\bm{w}\ssub{0}[4]$ \\ $\bm{w}\ssub{1}[4]$ \\ $\bm{w}\ssub{0}[5]$ \\ $\bm{w}\ssub{1}[5]$ \\ $\bm{w}\ssub{0}[6]$ \\ $\bm{w}\ssub{1}[6]$ \\ $\bm{w}\ssub{0}[7]$ \\ $\bm{w}\ssub{1}[7]$};
\node[draw, fill=green!60!black!10, align=center, font=\scriptsize, inner sep=4pt, right=0.4em of bk1s1] (bk1v1)
    {$\bm{w}\ssub{0}[0]$ \\ $\bm{w}\ssub{1}[0]$ \\ $\bm{w}\ssub{0}[1]$ \\ $\bm{w}\ssub{1}[1]$ \\ $\bm{w}\ssub{0}[2]$ \\ $\bm{w}\ssub{1}[2]$ \\ $\bm{w}\ssub{0}[3]$ \\ $\bm{w}\ssub{1}[3]$};
 
\foreach \i in {0,2,4,6} {
    \draw[red, thick, rounded corners=1pt] 
        ([yshift={-0.4em-\i*0.8em}, xshift=2pt]bk1s1.north west) 
        rectangle 
        ([yshift={-1.2em-\i*0.8em}, xshift=-2pt]bk1v1.north east);
}
 
\node[draw, fill=green!60!black!10, align=center, font=\scriptsize, inner sep=4pt, 
      below=2em of bk1s1] (bk1p)
    {$\bm{w}\ssub{0}[3]$ \\ $\bm{w}\ssub{1}[3]$ \\ $\bm{w}\ssub{0}[3]$ \\ $\bm{w}\ssub{1}[3]$ \\ $\bm{w}\ssub{0}[3]$ \\ $\bm{w}\ssub{1}[3]$ \\ $\bm{w}\ssub{0}[3]$ \\ $\bm{w}\ssub{1}[3]$};
\node[draw, fill=green!60!black!10, align=center, font=\scriptsize, inner sep=4pt, right=0.4em of bk1p] (bk1q)
    {$\bm{w}\ssub{1}[3]$ \\ $\bm{w}\ssub{0}[3]$ \\ $\bm{w}\ssub{1}[3]$ \\ $\bm{w}\ssub{0}[3]$ \\ $\bm{w}\ssub{1}[3]$ \\ $\bm{w}\ssub{0}[3]$ \\ $\bm{w}\ssub{1}[3]$ \\ $\bm{w}\ssub{0}[3]$};
 
\node[draw, align=center, font=\scriptsize, inner sep=4pt, right=2em of bk1v1] (bk2s1)
    {$\bm{w}\ssub{0}[0]$ \\ $\bm{w}\ssub{0}[1]$ \\ $\bm{w}\ssub{0}[2]$ \\ $\bm{w}\ssub{0}[3]$ \\ $\bm{w}\ssub{0}[4]$ \\ $\bm{w}\ssub{0}[5]$ \\ $\bm{w}\ssub{0}[6]$ \\ $\bm{w}\ssub{0}[7]$};
\node[draw, align=center, font=\scriptsize, inner sep=4pt, right=0.4em of bk2s1] (bk2v1)
    {$\bm{w}\ssub{1}[0]$ \\ $\bm{w}\ssub{1}[1]$ \\ $\bm{w}\ssub{1}[2]$ \\ $\bm{w}\ssub{1}[3]$ \\ $\bm{w}\ssub{1}[4]$ \\ $\bm{w}\ssub{1}[5]$ \\ $\bm{w}\ssub{1}[6]$ \\ $\bm{w}\ssub{1}[7]$};
 
\node[font=\scriptsize\itshape, above=0.3em of w0.north east] {Start};
\node[font=\scriptsize\itshape, above=0.3em of k1s1.north east] {round 1};
\node[font=\scriptsize\itshape, above=0.3em of k2s1.north east] {round 2};
\node[font=\scriptsize\itshape, above=0.3em of bk1s1.north east] {round 3};
\node[font=\scriptsize\itshape, above=0.3em of bk2s1.north east] {Final};
 
\node[font=\scriptsize, below=-0.2em of k1s1] {$\bm{u}$};
\node[font=\scriptsize, below=-0.2em of k1v1] {$\bm{v}$};
\node[font=\scriptsize, below=-0.2em of k2s1] {$\bm{u}$};
\node[font=\scriptsize, below=-0.2em of k2v1] {$\bm{v}$};
\node[font=\scriptsize, below=-0.2em of bk1s1] {$\bm{u}$};
\node[font=\scriptsize, below=-0.2em of bk1v1] {$\bm{v}$};
 
\node[font=\scriptsize, below=-0.2em of k1p] {$\bm{p}$};
\node[font=\scriptsize, below=-0.2em of k1q] {$\bm{q}$};
\node[font=\scriptsize, below=-0.2em of k2p] {$\bm{p}$};
\node[font=\scriptsize, below=-0.2em of k2q] {$\bm{q}$};
\node[font=\scriptsize, below=-0.2em of bk1p] {$\bm{p}$};
\node[font=\scriptsize, below=-0.2em of bk1q] {$\bm{q}$};
 
\draw[->, red, thick, >=Stealth, shorten >=3pt, shorten <=3pt] 
    (w1.east) -- node[above, font=\scriptsize\itshape, red] {merge} (k1s1.west);
 
\draw[->, red, thick, >=Stealth, shorten >=3pt, shorten <=3pt] 
    (k1v1.east) -- node[above, font=\scriptsize\itshape, red] {merge} (k2s1.west);
 
\draw[->, red, thick, >=Stealth, shorten >=3pt, shorten <=3pt] 
    (k2v1.east) -- node[above, font=\scriptsize\itshape, red] {merge} (bk1s1.west);
 
\draw[->, red, thick, >=Stealth, shorten >=3pt, shorten <=3pt] 
    (bk1v1.east) -- node[above, font=\scriptsize\itshape, red] {merge} (bk2s1.west);
 
\draw[->, green!60!black, thick, >=Stealth, shorten >=6pt, shorten <=6pt] 
    (k1v1.south) -- node[left, pos=0.35, font=\scriptsize\itshape, green!60!black] {copy} (k1p.north);
\draw[->, green!60!black, thick, >=Stealth, shorten >=3pt, shorten <=3pt] 
    (k1v1.south) -- node[right, pos=0.35, font=\scriptsize\itshape, green!60!black] {perm} (k1q.north);
 
\draw[->, green!60!black, thick, >=Stealth, shorten >=6pt, shorten <=6pt] 
    (k2v1.south) -- node[left, pos=0.35, font=\scriptsize\itshape, green!60!black] {perm} (k2p.north);
\draw[->, green!60!black, thick, >=Stealth, shorten >=3pt, shorten <=3pt] 
    (k2v1.south) -- node[right, pos=0.35, font=\scriptsize\itshape, green!60!black] {perm} (k2q.north);
 
\draw[->, green!60!black, thick, >=Stealth, shorten >=6pt, shorten <=6pt] 
    (bk1v1.south) -- node[left, pos=0.35, font=\scriptsize\itshape, green!60!black] {perm} (bk1p.north);
\draw[->, green!60!black, thick, >=Stealth, shorten >=3pt, shorten <=3pt] 
    (bk1v1.south) -- node[right, pos=0.35, font=\scriptsize\itshape, green!60!black] {perm} (bk1q.north);
 
\node[anchor=north, align=left, font=\scriptsize\itshape, inner sep=0pt] 
    at ([xshift=0.5em,yshift=-2.5em]$(bk2s1.south)!0.5!(bk2v1.south)$) {%
    $\bm{u}$: elements to be updated \\
    $\bm{v}$: elements not updated \\
    $\bm{p}$, $\bm{q}$: obtained by permuting $\bm{v}$ \\
    Each round updates $\bm{u}$ by \\
    $\bm{u} \mathrel{+}= \bm{c\ssub{p}} \odot \bm{p} + \bm{c\ssub{q}} \odot \bm{q}$ \\
    $\bm{c\ssub{p}}$, $\bm{c\ssub{q}}$: pre-computed vectors \\ 
    from $-\bm{C}$ and its powers \\
    };
 
\end{tikzpicture}%
}
\caption{Block-level realization of the Sklansky
construction for the $2 \times 2$ matrix
scan~\eqref{eq:recursive_doubling_equation} with $L = 8$;
the terminal blocks $\wbp\ssub{N-2}$ and $\wbp\ssub{N-1}$
are denoted $\bm{w}\ssub{0}$ and~$\bm{w}\ssub{1}$.  In
each round, merges assemble~$\bm{u}$ (blue, updated in
place) and~$\bm{v}$ (green); permutes
assemble~$\bm{p}$ and~$\bm{q}$.  The red rectangles mark
the elements that join~$\bm{u}$ in the following round,
their span doubling each round.  
E.g., in the first round,
$\bm{c}\ssub{p} = -[\bm{C}\ssub{00},\, \bm{C}\ssub{11},\,
\bm{C}\ssub{00},\, \bm{C}\ssub{11},\, \ldots]$ and
$\bm{c}\ssub{q} = -[\bm{C}\ssub{01},\, \bm{C}\ssub{10},\,
\bm{C}\ssub{01},\, \bm{C}\ssub{10},\, \ldots]$.}
\label{fig:block_realization_of_rd}

\end{figure*}
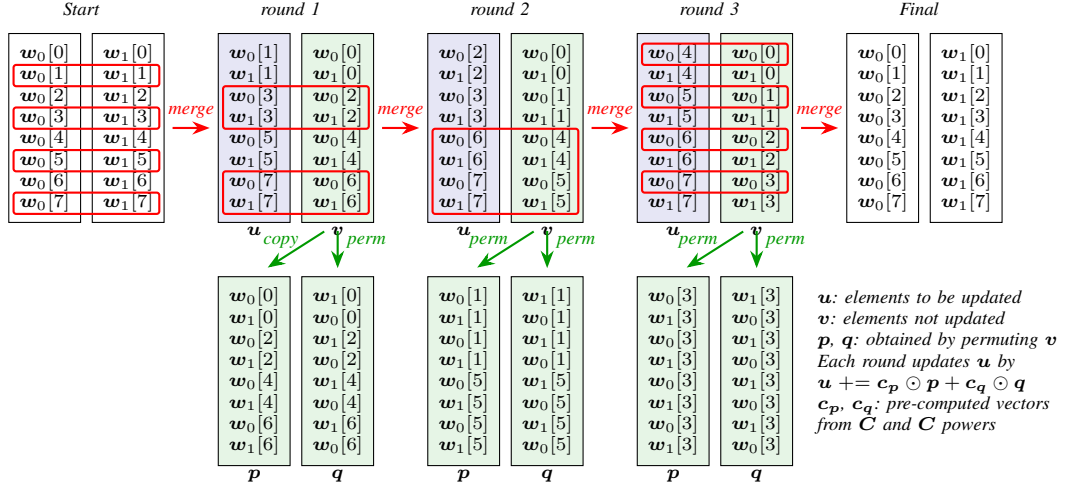

Guided by the signal flow
in~Figure~\ref{fig:recursive_doubling_diagram}, the
solution proceeds in $\log_2 L$ rounds.  The central
observation is that all $L/2$ matrix-vector products of a
round can be carried out as just two element-wise block
FMAs, because the matrix entries are known in advance: two
length-$L$ coefficient vectors $\bm{c}\ssub{p}$
and~$\bm{c}\ssub{q}$, pre-computed from the elements
of~$-\bm{C}$ and its powers, hold every entry in the SIMD lane
where it is needed.  At run time, block shuffles 
align the operands: two \emph{merges} split the $2L$
samples in~$\we[l]$ into a vector~$\bm{u}$, that is updated in
place, and a vector~$\bm{v}$, whose elements are
\emph{permuted} into the operand
vectors~$\bm{p}$ and~$\bm{q}$
(Figure~\ref{fig:block_realization_of_rd}).  Each round
then reduces to
\begin{equation}
\label{eq:block_rd_update}
\bm{u} = \bm{u} + \bm{c}\ssub{p} \odot \bm{p}
  + \bm{c}\ssub{q} \odot \bm{q},
\end{equation}
where $\odot$ denotes element-wise multiplication.

\subsubsection{PH and LU Algorithm Steps and Complexity}

The PH factorization solves the recursive 
stage~\eqref{eq:recursive} in three steps, which 
we now summarize together with their block-level 
operation counts (block FMAs and shuffles) and 
\emph{sequential depth}. 

\textit{Step~1: Particular solution.}
The forward substitution $\bm{P}\,\wg = \vg$
in~\eqref{eq:particular_and_homogeneous} computes a second-order
block recurrence
\begin{equation}
\label{eq:particular_recurrence}
\wbp\ssub{m} = \vbp\ssub{m}
  - a\ssub{1}\,\wbp\ssub{m-1}
  - a\ssub{2}\,\wbp\ssub{m-2},
  \quad m = 0, 1, \ldots, N{-}1,
\end{equation}
where $\wbp\ssub{-1} = \wbp\ssub{-2} = \bm{0}$. Thus, \eqref{eq:particular_recurrence}
requires $2$ block FMAs per block and $2N-3$ block FMAs in
total. Moreover, each block depends on its two immediate
predecessors, so the sequential depth of this step is~$N$.


\textit{Step~2: Terminal block solve.}
The two terminal blocks $\ybp\ssub{N-2}$ and
$\ybp\ssub{N-1}$ are computed by the block-level recursive
doubling described above, at a cost of
$2\log\ssub{2} L$ block FMAs and, counting the
$2\log\ssub{2} L - 1$ permutations and
$2\log\ssub{2} L + 2$ merges realized as intra-block
shuffles, $4\log\ssub{2} L + 1$ shuffles in total, with
sequential depth $\log\ssub{2} L$.

\textit{Step~3: Homogeneous correction.}
With the terminal blocks known, the remaining $N{-}2$ blocks
are recovered from~\eqref{eq:PH_forward}.
The two products $\SR\,\ybp\ssub{N-2}$ and
$\SR\,\ybp\ssub{N-1}$  each require a single block shuffle,
computed once and reused across all $N{-}2$ blocks. Each
correction then requires $2$ block FMAs, for a total of
$2(N{-}2)$ block FMAs and $2$ block shuffles. Since all
$N{-}2$ corrections are independent, this step has sequential
depth~$1$.

\begin{table}[t]
\centering
\caption{Per-stage complexity of PH-based multi-block filtering
for a signal block group of $NL$ samples.}
\label{tab:ph_complexity}
\scriptsize
\setlength{\tabcolsep}{4.8pt}
\begin{tabular}{lccc}
\toprule
Stage & Block FMAs & Shuffles & Seq.\ depth $\mathcal{O}(\cdot)$ \\
\midrule
Non-recursive & $2N+3$ & $2$ & $1$ \\
Particular solution & $2N-3$ & --- & $N$ \\
Recursive doubling & $2\log\ssub{2} L$
  & $4\log\ssub{2} L{+}1$ & $\log\ssub{2} L$ \\
Homogeneous correction & $2N-4$ & $2$ & $1$ \\
\midrule
Total & $6N{+}2\log\ssub{2} L{-}4$
  & $4\log\ssub{2} L{+}5$ & $N + \log\ssub{2} L$ \\
\midrule
Per sample & $\frac{6}{L}{+}\frac{2\log\ssub{2} L{-}4}{NL}$
  & $\frac{4\log\ssub{2} L{+}5}{NL}$ & $\frac{1}{L}{+} \frac{\log\ssub{2} L}{NL}$ \\
\bottomrule
\end{tabular}
\end{table}

\begin{table}[t]
\centering
\caption{Per-stage complexity of LU-based multi-block filtering
for a signal block group of $NL$ samples.}
\label{tab:lu_complexity}
\scriptsize
\setlength{\tabcolsep}{5pt}
\begin{tabular}{lccc}
\toprule
Stage & Block FMAs & Shuffles & Seq.\ depth $\mathcal{O}(\cdot)$ \\
\midrule
Non-recursive & $2N+3$ & $2$ & $1$ \\
Forward substitution & $2N-4+L$ & --- & $N{+}\log\ssub{2} L$ \\ 
Back substitution & $L+\log\ssub{2} L+2N-3$
  & $\log\ssub{2} L{+}2$ & $2\log\ssub{2} L$\\ 
\midrule
Total & $6N{+}2L{+}\log\ssub{2} L{-}4$
  & $\log\ssub{2} L{+}4$ & $N+3\log\ssub{2} L$ \\
\midrule
Per sample & $\frac{6}{L} {+}\frac{2}{N}
  {+}\frac{\log\ssub{2} L{-}4}{NL}$
  & $\frac{\log\ssub{2} L{+}4}{NL}$ & $\frac{1}{L}{+} \frac{3\log\ssub{2} L}{NL}$ \\
\bottomrule
\end{tabular}
\end{table}

Table~\ref{tab:ph_complexity} summarizes the per-stage costs of PH factorization.
Including the non-recursive
stage~\eqref{eq:feedforward_block_uniform}, the full PH-based
multi-block filtering algorithm requires approximately
$6N {+} 2\log\ssub{2} L$ block FMAs and $4\log\ssub{2} L$ block shuffles
per signal block group of $NL$ samples.

Table~\ref{tab:lu_complexity} summarizes the corresponding
costs for the LU factorization.
The $6N$ block FMAs are common to both factorizations.
The difference lies in solving the terminal blocks.
With LU factorization, dense matrices $\bm{K}$ and
$\bm{J}$ cost a total of $2L$~block FMAs for the last
block in forward and back substitution.  
The second-to-last
block, coupled through $(\bm{I}{+}\bm{E})$, reduces to a
first-order scalar recurrence across the $L$ samples within
the block, which can be solved by recursive doubling at a
cost of $\log\ssub{2} L$~block FMAs and $\log\ssub{2} L$~shuffles.
By contrast, PH factorization couples both terminal
blocks into a single first-order, $2 \times 2$ recurrence of length~$L$,
solved by recursive doubling at a cost of only
$2\log\ssub{2} L$~block FMAs.

Tables~\ref{tab:ph_complexity} and~\ref{tab:lu_complexity}
also illustrate the sequential dependency depth.  
In both factorizations, the depth is dominated by
the particular solution (PH) or the forward substitution
(LU), each imposing $\mathcal{O}(N)$ sequential block dependencies.


\subsection{Cyclic Reduction}
\label{sec:cyclic_reduction}

For cyclic reduction, the number of blocks $N$ is assumed to be a power of 2.

The block tridiagonal structure of~$\Ag$
in~\eqref{eq:multi_block_A} reveals that each
block~$\ybp\ssub{m}$ is coupled to its two neighboring
blocks through the scalar multiples~$a\ssub{1}\bm{I}$
and~$a\ssub{2}\bm{I}$.  This nearest-neighbor coupling
suggests that alternating blocks can be systematically
eliminated, reducing the problem size by half at each
level.

\subsubsection{Reduction}

To begin, the $N$ equations~\eqref{eq:recursive} for the block recursive stage are partitioned
into an
even-indexed group of blocks
$\ybp\ssub{0}, \ybp\ssub{2}, \ldots, \ybp\ssub{N-2}$ and
an odd-indexed group
$\ybp\ssub{1}, \ybp\ssub{3}, \ldots, \ybp\ssub{N-1}$,   
\begin{equation*}
\label{eq:cr_permuted_system}
\scalebox{0.82}{$
\left[
\begin{NiceArray}{cccc:cccc}
 \bm{I} &   &   & a\ssub{2}\SR  &   &   &   & a\ssub{1}\SR \\
 a\ssub{2}\bm{I} &  \bm{I} &   &   & a\ssub{1}\bm{I}  &   &   &   \\
  & \ddots  & \ddots  &   &   & \ddots  &   &   \\
  &   & a\ssub{2}\bm{I}  & \bm{I}  &   &   & a\ssub{1}\bm{I}  &   \\
\hdashline
 a\ssub{1}\bm{I} &   &   &   & \bm{I}  &   &   & a\ssub{2}\SR  \\
  & a\ssub{1}\bm{I}  &   &   & a\ssub{2}\bm{I}  & \bm{I}  &   &   \\
  &   &  \ddots &   &   & \ddots  & \ddots  &   \\
  &   &   & a\ssub{1}\bm{I} &   &   & a\ssub{2}\bm{I}  & \bm{I} 
\end{NiceArray}
\right]
\begin{bmatrix}
\ybp\ssub{0} \\ \ybp\ssub{2} \\ \vdots \\ \ybp\ssub{N-2} \\
\hdashline
\ybp\ssub{1} \\ \ybp\ssub{3} \\ \vdots \\ \ybp\ssub{N-1}
\end{bmatrix}
=
\begin{bmatrix}
\vbp\ssub{0} \\ \vbp\ssub{2} \\ \vdots \\ \vbp\ssub{N-2} \\
\hdashline
\vbp\ssub{1} \\ \vbp\ssub{3} \\ \vdots \\ \vbp\ssub{N-1} \\
\end{bmatrix}.
$}
\end{equation*}
%
The even-indexed blocks are eliminated from the system by
Gaussian elimination in two steps.  The elimination is the
first of $\log_2 N$ identical reduction levels; the
superscript on the right-hand side counts these levels,
$\vbp\ssub{n}^{(i)}$ denoting the block after the updates
of level~$i{-}1$, with $\vbp\ssub{n}^{(0)} = \vbp\ssub{n}$.
In the first step, even-indexed rows~$2n$ with $n>0$  are modified
by subtracting $a\ssub{2}/a\ssub{1}$ times
row~$2n{-}1$.  
For row $0$, $a\ssub{2}/a\ssub{1} \bm{R}$ times row~$N{-}1$ is subtracted.
Correspondingly, the even-indexed right-hand sides are
updated by
\begin{equation}
    \label{eq:CR_update_even}
    \begin{aligned}
        \vbp\ssub{2n}^{(1)} & = \vbp\ssub{2n}^{(0)} - \frac{a\ssub{2}}{a\ssub{1}} \vbp\ssub{2n-1}^{(0)} \text{, $n=1,\ldots,N/2-1$}\\
        \vbp\ssub{0}^{(1)} & = \vbp\ssub{0}^{(0)} - \frac{a\ssub{2}}{a\ssub{1}} \bm{R}\vbp\ssub{N-1}^{(0)}.
    \end{aligned}
\end{equation}
In the second step, each odd-indexed row~$2n{+}1$ is updated by
subtracting~$a\ssub{1}$ times the modified even-indexed
row~$2n$.
The right-hand side is updated by
\begin{equation}
    \label{eq:CR_update_odd}
    \vbp\ssub{2n+1}^{(1)} = \vbp\ssub{2n+1}^{(0)} - a\ssub{1} \vbp\ssub{2n}^{(1)} \text{, $n=0,\ldots,N/2-1$.}
\end{equation}
%
The odd-indexed blocks form a reduced system of~$N/2$
blocks with updated coefficients.  Denoting the system at
level~$i$ by $\Ag\ssub{CR}^{(i)}\,\yg^{(i)} = \vg^{(i)}$,
with $\Ag\ssub{CR}^{(0)} = \Ag$, the two elimination steps
transform the partitioned matrix as
\begin{equation}
\label{eq:cr_general_structure}
\scalebox{0.82}{$
\Ag\ssub{CR}^{(i)}
\xrightarrow{\text{elim.}}
\left[
  \begin{NiceArray}{cccc:cccc}
    \bm{I} & & & & & & d^{(i+1)}\SR  & c^{(i+1)}\SR  \\
    & \bm{I} & & & c^{(i+1)}\bm{I} & & & d^{(i+1)}\SR \\
    & & \ddots & & d^{(i+1)}\bm{I} & \ddots & & \\
    & & & \bm{I} & & \ddots & c^{(i+1)}\bm{I} & \\
    \hdashline
    & & & & \bm{I} & & a\ssub{2}^{(i+1)}\SR & a\ssub{1}^{(i+1)}\SR \\
    & & & & a\ssub{1}^{(i+1)}\bm{I} & \bm{I} & & a\ssub{2}^{(i+1)}\SR \\
    & & & & a\ssub{2}^{(i+1)}\bm{I} & \ddots & \ddots & \\
    & & & & & \ddots & a\ssub{1}^{(i+1)}\bm{I} & \bm{I} \\
    \CodeAfter
      \UnderBrace{8-5}{8-8}{\Ag\ssub{CR}^{(i+1)}}
  \end{NiceArray}
  \right]
$}
\vspace{0.4cm}
\end{equation}
%
for $i = 0, \ldots, \log\ssub{2} N {-} 3$.  
At each
level, the right-hand side is transformed
from~$\vbp\ssub{n}^{(i)}$ to~$\vbp\ssub{n}^{(i+1)}$ by
updates of the form~\eqref{eq:CR_update_even}
and~\eqref{eq:CR_update_odd}, with $a\ssub{1}^{(i)}$,
$a\ssub{2}^{(i)}$ in place of $a\ssub{1}$, $a\ssub{2}$.
The upper half contains the blocks eliminated at the
current level, their right-hand sides frozen
at~$\vbp\ssub{n}^{(i+1)}$.
Once the blocks of the lower
half are computed, the upper-half blocks are recovered by
back substitution using~$c^{(i+1)}$ and~$d^{(i+1)}$.  
The
lower-right quadrant forms the  reduced
system~$\Ag\ssub{CR}^{(i+1)}$, which operates on half the
number of blocks and retains the same block structure with
updated coefficients~$a\ssub{1}^{(i+1)}$
and~$a\ssub{2}^{(i+1)}$.  The coefficients are initialized
as $a\ssub{1}^{(0)} = a\ssub{1}$,
$a\ssub{2}^{(0)} = a\ssub{2}$, and updated at each level
by
%
%
\begin{equation}
\label{eq:cr_coeff_update}
\begin{aligned}
d^{(i+1)} &= -\frac{a\ssub{2}^{(i)}a\ssub{2}^{(i)}}{a\ssub{1}^{(i)}}, &\quad
c^{(i+1)} &= a\ssub{1}^{(i)}-\frac{a\ssub{2}^{(i)}}{a\ssub{1}^{(i)}}, \\
a\ssub{2}^{(i+1)} &= a\ssub{2}^{(i)}a\ssub{2}^{(i)}, &\quad
a\ssub{1}^{(i+1)} &= 2a\ssub{2}^{(i)}-a\ssub{1}^{(i)}a\ssub{1}^{(i)}.
\end{aligned}
\end{equation}
%
These coefficients depend only on the filter
parameters~$a\ssub{1}$ and~$a\ssub{2}$ and can therefore
be pre-computed before filtering begins.
Note that for biquads with complex-conjugate poles, 
$r^{(i)} e^{\pm j \theta^{(i)}}$, at level~$i$ 
(i.e.,
$a\ssub{2}^{(i)} = (r^{(i)})^2$
and~$a\ssub{1}^{(i)} = -2r^{(i)} \cos(\theta^{(i)})$),
a simple recursion for the poles follows from~\eqref{eq:cr_coeff_update},
\begin{equation}
    \label{eq:recursion_poles}
    r^{(i+1)} = (r^{(i)})^2 \text{ and }
    \theta^{(i+1)} = 2\theta^{(i)}.
\end{equation}
For biquads with real-valued poles at $r\ssub{1}^{(i)}$
and $r\ssub{2}^{(i)}$, the next level poles are given by
$(r\ssub{1}^{(i)})^2$ and $(r\ssub{2}^{(i)})^2$.

As the system shrinks, the last two levels deviate from
the general pattern~\eqref{eq:cr_general_structure}.
At level~$i_2 = \log\ssub{2} N {-} 2$, only four blocks
remain, with each quadrant reducing to a single
$2L \times 2L$ block,
\begin{equation*}
\label{eq:cr_four_blocks}
\Ag\ssub{CR}^{(i_2)}
\;\xrightarrow{\;\text{elim.}\;}\;
\left[
\begin{NiceArray}{cc:cc}
  \bm{I} &          & d^{(i_2+1)}\SR      & c^{(i_2+1)}\SR      \\
         & \bm{I}   & c^{(i_2+1)}\bm{I}   & d^{(i_2+1)}\SR      \\
  \hdashline
         &          & \bm{I}+a\ssub{2}^{(i_2+1)}\SR & a\ssub{1}^{(i_2+1)}\SR    \\
         &          & a\ssub{1}^{(i_2+1)}\bm{I}     & \bm{I}+a\ssub{2}^{(i_2+1)}\SR \\
\end{NiceArray}
\right].
\end{equation*}
At the final level~$i_1 = \log\ssub{2} N {-} 1$, the
system reduces to a single $2L \times 2L$ system,
\begin{equation}
\label{eq:cr_two_blocks}
\Ag\ssub{CR}^{(i_1)}
\;\xrightarrow{\;\text{elim.}\;}\;
\left[
\begin{array}{cc}
\bm{I}  & c^{(i_1+1)}\SR + d^{(i_1+1)}\SR^{2} \\
 \bm{0} & \bm{I} + a\ssub{1}^{(i_1+1)}\SR + a\ssub{2}^{(i_1+1)}\SR^{2}
\end{array}
\right],
\end{equation}
where~$\SR^2$, the square of the shift matrix, has ones
on the second sub-diagonal.  After this final elimination,
the block~$\ybp\ssub{N-1}$ decouples and is solved
independently.
The companion block~$\ybp\ssub{N/2-1}$ in~\eqref{eq:cr_two_blocks} is
then recovered by back substitution.

\textbf{What if $a\ssub{1}^{(i)} \approx 0$?}
The coefficient update~\eqref{eq:cr_coeff_update} 
and the elimination~\eqref{eq:CR_update_even}
contain
the only divisions of the algorithm: the reduction
multiplier~$a\ssub{2}^{(i)}/a\ssub{1}^{(i)}$, the
even-indexed updates~$\vbp\ssub{2n}^{(i+1)}$, and the
recovery coefficients~$c^{(i+1)}$ and~$d^{(i+1)}$ all
scale as~$1/a\ssub{1}^{(i)}$.  
The elimination at
level~$i$ therefore becomes ill-conditioned
when~$a\ssub{1}^{(i)}$ approaches zero.
The blocks recovered
at that level emerge as nearly canceling differences of
terms roughly~$|a\ssub{2}^{(i)}/a\ssub{1}^{(i)}|$ times
larger than the result, and their rounding error
therefore grows in proportion to this multiplier.

The remedy replaces the reduction at the affected
level~$i^{\ast}$ by a division-free row combination; we
refer to this as \emph{splitting} the level.  
In $z$-plane terms, multiplying the denominator
$1 + a\ssub{1}^{(i^{\ast})}z^{-1} + a\ssub{2}^{(i^{\ast})}z^{-2}$ by its
mirror $1 - a\ssub{1}^{(i^{\ast})}z^{-1} + a\ssub{2}^{(i^{\ast})}z^{-2}$
cancels the odd powers of~$z^{-1}$ and squares the
poles -- the scattered look-ahead transformation
of~\cite{Parhi_89}.  
In the block-matrix framework, the
mirror polynomial is the Toeplitz matrix~$\bm{M}$ obtained
from~$\Ag\ssub{CR}^{(i^{\ast})}$ by negating its middle
coefficient~$a\ssub{1}^{(i^{\ast})}$.
The split transforms the
level-$i^{\ast}$ system
$\Ag\ssub{CR}^{(i^{\ast})}\,\yg^{(i^{\ast})} =
\vg^{(i^{\ast})}$ into
$\bm{M}\Ag\ssub{CR}^{(i^{\ast})}\,\yg^{(i^{\ast})} =
\bm{M}\vg^{(i^{\ast})}$, where
\begin{equation}
\label{eq:split_MA}
\bm{M}\Ag\ssub{CR}^{(i^{\ast})} =
\scalebox{0.8}{$
\begin{bmatrix}
\bm{I} & & & a\ssub{2}'\SR & \bm{0} & a\ssub{1}'\SR & \bm{0} \\
\bm{0} & \bm{I} & & & a\ssub{2}'\SR & \bm{0} & a\ssub{1}'\SR  \\
a\ssub{1}'\bm{I} & \bm{0} & \bm{I} & & & a\ssub{2}'\SR &  \bm{0}\\
\bm{0} & a\ssub{1}'\bm{I} & \bm{0} & \bm{I} & & & a\ssub{2}'\SR  \\
a\ssub{2}'\bm{I} & \bm{0} & a\ssub{1}'\bm{I} & \bm{0} & \bm{I} & &  \\
& \ddots & \ddots & \ddots & \ddots & \ddots &  \\
& & a\ssub{2}'\bm{I} & \bm{0} & a\ssub{1}'\bm{I} & \bm{0} & \bm{I}
\end{bmatrix}
$}
\end{equation}
with $a\ssub{1}' = 2a\ssub{2}^{(i^{\ast})} - a\ssub{1}^{(i^{\ast})}a\ssub{1}^{(i^{\ast})}$
and~$a\ssub{2}' = a\ssub{2}^{(i^{\ast})}a\ssub{2}^{(i^{\ast})}$, 
i.e., the same as the level-$(i^{\ast}{+}1)$ coefficient updates
in~\eqref{eq:cr_coeff_update}.  
The odd lags have cancelled; 
each block couples only to blocks two and four
positions away, i.e., to blocks of its own parity.
Partitioning the even- from the odd-indexed blocks 
therefore yields two
independent half-size systems of the same form, on which
the reduction proceeds separately.

Small values of $a\ssub{1}^{(i)}$ occur whenever the $i$-th level poles are near
$\theta^{(i)} \approx \pm \frac{\pi}{2}$.
Because of the pole recursion~\eqref{eq:recursion_poles},
this may occur at any level and may even occur multiple times.
Hence, the above remedy is applied any time a small value of $a\ssub{1}^{(i)}$
is encountered.
As shown in Section~\ref{sec:accuracy}, numerical problems are averted when splitting is used whenever 
$|a\ssub{1}^{(i)}| < \frac{1}{4}|a\ssub{2}^{(i)}|$.


\subsubsection{Back Substitution}


The back substitution proceeds in reverse order of the
reduction, from level~$i_1 = \log\ssub{2} N {-} 1$ down
to~$i = 0$.  Referring to the general
structure~\eqref{eq:cr_general_structure}, the lower-right
quadrant at each level is the reduced system whose blocks
have already been solved at the deeper levels; the upper
half provides the equations from which the eliminated
blocks are recovered, each combining the frozen
right-hand side~$\vbp\ssub{n}^{(i+1)}$ with the
coefficients~$c^{(i+1)}$ and~$d^{(i+1)}$.

At the final reduction level~$i_1 = \log\ssub{2} N {-} 1$, the
system reduces to the two-block triangular system
in~\eqref{eq:cr_two_blocks}.
The last
block~$\ybp\ssub{N-1}$ is obtained from
\begin{equation}\label{eq:cr_terminal_block}
    \bigl(\bm{I} + a\ssub{1}^{(i_1+1)}\,\SR
    + a\ssub{2}^{(i_1+1)}\,\SR^2\bigr)\,\ybp\ssub{N-1}
  = \vbp\ssub{N-1}^{(i_1+1)},
\end{equation}
where~$\vbp\ssub{N-1}^{(i_1+1)}$ has accumulated the
updates of all~$\log\ssub{2} N$ reduction levels.
Since
$(\bm{I} + a\ssub{1}^{(i_1+1)}\,\SR + a\ssub{2}^{(i_1+1)}\,\SR^2)$
is a lower-triangular Toeplitz matrix with three non-zero
bands,~\eqref{eq:cr_terminal_block} reduces to block
filtering~\eqref{eq:block_filtering_equation_An} and can be
solved with~$L$~block
FMAs.\footnote{Alternatively, \eqref{eq:cr_terminal_block}
can be solved by recursive doubling.
For vector length $L\leq 8$ this is less efficient.}

The upper companion block~$\ybp\ssub{N/2-1}$ is then
recovered from~\eqref{eq:cr_two_blocks}
once~$\ybp\ssub{N-1}$ is known,
\begin{equation}\label{eq:cr_companion_block}
  \ybp\ssub{N/2-1}
    = \vbp\ssub{N/2-1}^{(i_1+1)}
      - c^{(i_1+1)}\,\SR\,\ybp\ssub{N-1}
      - d^{(i_1+1)}\,\SR^2\,\ybp\ssub{N-1}.
\end{equation}

At each subsequent level from~$i = \log\ssub{2} N - 2$ down
to~$i = 0$, the eliminated blocks are recovered from the
blocks already solved at the deeper levels, each requiring
$2$~block FMAs with the coefficients~$c^{(i+1)}$
and~$d^{(i+1)}$.
At the final
level~$i = 0$, all even-indexed blocks 
$\ybp\ssub{2m}$, $m=0, \ldots, \frac{N}{2}{-}1$
are recovered as
\begin{equation}
\label{eq:cr_backsub_final}
\begin{aligned}
  \ybp\ssub{0}
    &= \vbp\ssub{0}^{(1)}
      - d^{(1)}\,\SR\,\ybp\ssub{N-3}
      - c^{(1)}\,\SR\,\ybp\ssub{N-1}
      , \\
  \ybp\ssub{2}
    &= \vbp\ssub{2}^{(1)}
      - c^{(1)}\,\ybp\ssub{1}
      - d^{(1)}\,\SR\,\ybp\ssub{N-1}, \\
  \ybp\ssub{2m}
    &= \vbp\ssub{2m}^{(1)}
      - c^{(1)}\,\ybp\ssub{2m-1}
      - d^{(1)}\,\ybp\ssub{2m-3},
\end{aligned}
\end{equation}
where the first two blocks ($m = 0,\,1$) involve the shift
matrix~$\SR$ in place of~$\bm{I}$, as the block indices
wrap around modulo~$N$.
Note that $\SR\,\ybp\ssub{N-1}$ was already computed, but
$\SR\,\ybp\ssub{N-3}$ requires one more shuffle.

\subsubsection{CR Algorithm Steps and Complexity}

The cyclic reduction solves the recursive
stage~\eqref{eq:recursive} in three steps, which we now
summarize together with their block-level costs.

\textit{Step~1: Reduction.}
Starting from the full system of~$N$ blocks, each reduction
level eliminates alternating blocks to produce a system of
half the size with updated coefficients, as described above.
At each level, all block updates~\eqref{eq:CR_update_even} and~\eqref{eq:CR_update_odd} are independent, and each
requires one block FMA plus one shuffle for the boundary
block.  Because each level halves the system size over
$\log\ssub{2} N$ levels, the total cost is $2N{-}2$ block
FMAs and $\log\ssub{2} N$ shuffles, with sequential
depth~$\log\ssub{2} N$.

\textit{Step~2: Terminal block filtering.}
At the deepest reduction level, the system reduces to a
single block governed
by~\eqref{eq:cr_terminal_block}, a banded lower triangular
Toeplitz system with three bands.  This is solved by block
filtering at a cost of~$L$ block FMAs, with sequential
depth~$\log\ssub{2}L$.

\textit{Step~3: Back substitution.}
The substitution proceeds from the deepest level back to
level~$i = 0$, recovering twice as many blocks at each
level.  All recoveries within a given level are
independent, so the sequential depth is~$\log\ssub{2} N$.
Each recovered block requires two block FMAs.  
At the
deepest level, recovering the companion
block~\eqref{eq:cr_companion_block} requires $2$~shuffles:
one for~$\SR^2\,\ybp\ssub{N-1}$ and one
for~$\SR\,\ybp\ssub{N-1}$.  The
result~$\SR\,\ybp\ssub{N-1}$ is then reused at all
subsequent levels, as shown
in~\eqref{eq:cr_backsub_final}.  
At each of the
remaining~$\log\ssub{2} N$ levels, the boundary block on the first row, 
due to~$d^{(i)}\SR$, requires one
additional shuffle per level.  By the same 
argument as in Step~1, the total cost is $2N{-}2$ block
FMAs and $\log\ssub{2} N + 2$ shuffles.

Table~\ref{tab:cr_complexity} summarizes the per-stage
costs.  Including the non-recursive
stage~\eqref{eq:feedforward_block_uniform}, the full
CR-based multi-block filtering algorithm requires
approximately $6N + L$ block FMAs and
$2\log\ssub{2} N$ block shuffles per signal block group
of~$NL$ samples.  The block FMA total is comparable to that
of PH factorization, as both are dominated by the~$6N$
term.
However, the sequential depth is reduced
from~$\mathcal{O}(N)$ to~$\mathcal{O}(\log\ssub{2} N)$ relative to PH factorization.

When a section requires the split as
in~\eqref{eq:split_MA}, these counts change only
marginally. 
The exact number of additional operations depends on the level $i^*$
where the split happens.
Like all level coefficients, the matrix $\bm{M}$ depends
only on the pre-computed
pair~$(a\ssub{1}^{(i^{\ast})}, a\ssub{2}^{(i^{\ast})})$, 
so the split is
fixed before filtering begins.  
The reduction and the back
substitution proceed as before, on two subsystems instead
of one, at an essentially unchanged block-FMA count. 
The
additional work is the application of~$\bm{M}$ to the
right-hand side ($2N/2^{i^*}$ FMAs), the second terminal solve, 
one per subsystem ($L$ FMAs), 
and $(\log_2N-1-i^*)$ additional shuffles during back substitution.  
The sequential depth is unchanged, since the two
subsystems proceed in parallel through reduction, terminal
solve, and back substitution.  The split therefore removes
the hazardous divisions at a small arithmetic cost, with
the sequential depth intact.

\begin{table}[t]
\centering
\caption{Per-stage complexity of CR-based multi-block
filtering for a signal block group of $NL$ samples.}
\label{tab:cr_complexity}
\scriptsize
\setlength{\tabcolsep}{5.5pt}
\begin{tabular}{lccc}
\toprule
Stage & Block FMAs & Shuffles & Seq.\ depth $\mathcal{O}(\cdot)$ \\
\midrule
Non-recursive
  & $2N{+}3$ & $2$ & $1$ \\
Reduction
  & $2N{-}2$ & $\log\ssub{2} N$ & $\log\ssub{2} N$ \\
Terminal block filtering
  & $L$ & --- & $\log\ssub{2} L$ \\
Back substitution
  & $2N{-}2$ & $\log\ssub{2} N{+}2$ & $\log\ssub{2} N$ \\
\midrule
Total
  & $6N{+}L{-}1$
  & $2\log\ssub{2} N{+}4$ & $2\log\ssub{2} N{+}\log\ssub{2} L$ \\
\midrule
Per sample
  & $\frac{6}{L}{+}\frac{1}{N}{-}\frac{1}{NL}$
  & $\frac{2\log\ssub{2} N{+}4}{NL}$ & $\frac{2\log\ssub{2} N{+}\log\ssub{2} L}{NL}$\\
\bottomrule
\end{tabular}
\end{table}

\section{Realization of High-Order Systems}
\label{sec:permutation}

\noindent The preceding sections developed the multi-block filtering
algorithms for a single second-order section.  Before
extending the framework to general high-order systems
realized as cascaded biquads, we address a remaining
component: the block permutation, which is the enabling
operation that maps the input $NL$ samples into the stride-$N$
layout required by~\eqref{eq:multi_block_A}.

\subsection{Block Permutation}
\label{sec:block_permutation}

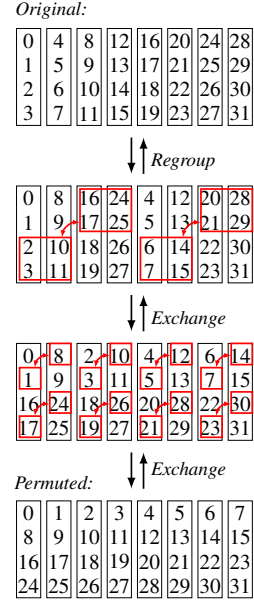
\begin{figure}
\centering
    \scalebox{0.8}{$
    \begin{tikzpicture}[scale=0.2]

    \draw (0,0) rectangle (2,8);
    \node at (1,1) {3};
    \node at (1,3) {2};
    \node at (1,5) {1};
    \node at (1,7) {0};

    \begin{scope}[xshift=2.5cm]
        \draw (0,0) rectangle (2,8);
        \node at (1,1) {7};
        \node at (1,3) {6};
        \node at (1,5) {5};
        \node at (1,7) {4};
    \end{scope}

    \begin{scope}[xshift=5cm]
        \draw (0,0) rectangle (2,8);
        \node at (1,1) {11};
        \node at (1,3) {10};
        \node at (1,5) {9};
        \node at (1,7) {8};
    \end{scope}

    \begin{scope}[xshift=7.5cm]
        \draw (0,0) rectangle (2,8);
        \node at (1,1) {15};
        \node at (1,3) {14};
        \node at (1,5) {13};
        \node at (1,7) {12};
    \end{scope}

    \begin{scope}[xshift=10cm]
        \draw (0,0) rectangle (2,8);
        \node at (1,1) {19};
        \node at (1,3) {18};
        \node at (1,5) {17};
        \node at (1,7) {16};
    \end{scope}

    \begin{scope}[xshift=12.5cm]
        \draw (0,0) rectangle (2,8);
        \node at (1,1) {23};
        \node at (1,3) {22};
        \node at (1,5) {21};
        \node at (1,7) {20};
    \end{scope}

    \begin{scope}[xshift=15cm]
        \draw (0,0) rectangle (2,8);
        \node at (1,1) {27};
        \node at (1,3) {26};
        \node at (1,5) {25};
        \node at (1,7) {24};
    \end{scope}

    \begin{scope}[xshift=17.5cm]
        \draw (0,0) rectangle (2,8);
        \node at (1,1) {31};
        \node at (1,3) {30};
        \node at (1,5) {29};
        \node at (1,7) {28};
    \end{scope}

    \node[left,font=\small\itshape] at (6.5,9.5) {Original:};

    \draw[-latex, thick] (9.5,-1) -- (9.5,-4);
    \node[right,font=\small\itshape] at (10.5,-3) {Regroup};
    \draw[-latex, thick] (10.5,-4) -- (10.5,-1);

    \begin{scope}[yshift=-13cm]
        \draw (0,0) rectangle (2,8);
        \node at (1,1) {3};
        \node at (1,3) {2};
        \node at (1,5) {1};
        \node at (1,7) {0};

        \begin{scope}[xshift=2.5cm]
            \draw (0,0) rectangle (2,8);
            \node at (1,1) {11};
            \node at (1,3) {10};
            \node at (1,5) {9};
            \node at (1,7) {8};
        \end{scope}

        \begin{scope}[xshift=5cm]
            \draw (0,0) rectangle (2,8);
            \node at (1,1) {19};
            \node at (1,3) {18};
            \node at (1,5) {17};
            \node at (1,7) {16};
        \end{scope}

        \begin{scope}[xshift=7.5cm]
            \draw (0,0) rectangle (2,8);
            \node at (1,1) {27};
            \node at (1,3) {26};
            \node at (1,5) {25};
            \node at (1,7) {24};
        \end{scope}

        \begin{scope}[xshift=10cm]
            \draw (0,0) rectangle (2,8);
            \node at (1,1) {7};
            \node at (1,3) {6};
            \node at (1,5) {5};
            \node at (1,7) {4};
        \end{scope}

        \begin{scope}[xshift=12.5cm]
            \draw (0,0) rectangle (2,8);
            \node at (1,1) {15};
            \node at (1,3) {14};
            \node at (1,5) {13};
            \node at (1,7) {12};
        \end{scope}

        \begin{scope}[xshift=15cm]
            \draw (0,0) rectangle (2,8);
            \node at (1,1) {23};
            \node at (1,3) {22};
            \node at (1,5) {21};
            \node at (1,7) {20};
        \end{scope}

        \begin{scope}[xshift=17.5cm]
            \draw (0,0) rectangle (2,8);
            \node at (1,1) {31};
            \node at (1,3) {30};
            \node at (1,5) {29};
            \node at (1,7) {28};
        \end{scope}

        \draw[red, thick] (0.25,0.25) rectangle (4.5,3.75);
        \draw[red, thick] (5.25,4.25) rectangle (9.5,7.75);
        \draw[<->, red, thick,>={Latex[scale=0.5]}](3.75,3.75) to[bend left=40] (5.25,5);

        \draw[red, thick] (0.25+10,0.25) rectangle (4.5+10,3.75);
        \draw[red, thick] (5.25+10,4.25) rectangle (9.5+10,7.75);
        \draw[<->, red, thick,>={Latex[scale=0.5]}](3.75+10,3.75) to[bend left=40] (5.25+10,5);

        \draw[-latex, thick] (9.5,-1) -- (9.5,-4);
        \node[right,font=\small\itshape] at (10.5,-3) {Exchange};
        \draw[-latex, thick] (10.5,-4) -- (10.5,-1);
    \end{scope}

    \begin{scope}[yshift=-26cm]
        \draw (0,0) rectangle (2,8);
        \node at (1,1) {17};
        \node at (1,3) {16};
        \node at (1,5) {1};
        \node at (1,7) {0};

        \begin{scope}[xshift=2.5cm]
            \draw (0,0) rectangle (2,8);
            \node at (1,1) {25};
            \node at (1,3) {24};
            \node at (1,5) {9};
            \node at (1,7) {8};
        \end{scope}

        \begin{scope}[xshift=5cm]
            \draw (0,0) rectangle (2,8);
            \node at (1,1) {19};
            \node at (1,3) {18};
            \node at (1,5) {3};
            \node at (1,7) {2};
        \end{scope}

        \begin{scope}[xshift=7.5cm]
            \draw (0,0) rectangle (2,8);
            \node at (1,1) {27};
            \node at (1,3) {26};
            \node at (1,5) {11};
            \node at (1,7) {10};
        \end{scope}

        \begin{scope}[xshift=10cm]
            \draw (0,0) rectangle (2,8);
            \node at (1,1) {21};
            \node at (1,3) {20};
            \node at (1,5) {5};
            \node at (1,7) {4};
        \end{scope}

        \begin{scope}[xshift=12.5cm]
            \draw (0,0) rectangle (2,8);
            \node at (1,1) {29};
            \node at (1,3) {28};
            \node at (1,5) {13};
            \node at (1,7) {12};
        \end{scope}

        \begin{scope}[xshift=15cm]
            \draw (0,0) rectangle (2,8);
            \node at (1,1) {23};
            \node at (1,3) {22};
            \node at (1,5) {7};
            \node at (1,7) {6};
        \end{scope}

        \begin{scope}[xshift=17.5cm]
            \draw (0,0) rectangle (2,8);
            \node at (1,1) {31};
            \node at (1,3) {30};
            \node at (1,5) {15};
            \node at (1,7) {14};
        \end{scope}

        \draw[red, thick] (0.25,4.25) rectangle (2,6);
        \draw[red, thick] (5.25,4.25) rectangle (7,6);
        \draw[red, thick] (10.25,4.25) rectangle (12,6);
        \draw[red, thick] (15.25,4.25) rectangle (17,6);

        \draw[red, thick] (2.75,6.25) rectangle (4.5,8);
        \draw[red, thick] (7.75,6.25) rectangle (9.5,8);
        \draw[red, thick] (12.75,6.25) rectangle (14.5,8);
        \draw[red, thick] (17.75,6.25) rectangle (19.5,8);

        \draw[red, thick] (0.25,0.25) rectangle (2,2);
        \draw[red, thick] (5.25,0.25) rectangle (7,2);
        \draw[red, thick] (10.25,0.25) rectangle (12,2);
        \draw[red, thick] (15.25,0.25) rectangle (17,2);

        \draw[red, thick] (2.75,2.25) rectangle (4.5,4);
        \draw[red, thick] (7.75,2.25) rectangle (9.5,4);
        \draw[red, thick] (12.75,2.25) rectangle (14.5,4);
        \draw[red, thick] (17.75,2.25) rectangle (19.5,4);

        \draw[<->, red, thick,>={Latex[scale=0.5]}](1.5,2) to[bend left=40] (2.75,3);
        \draw[<->, red, thick,>={Latex[scale=0.5]}](6.5,2) to[bend left=40] (7.75,3);
        \draw[<->, red, thick,>={Latex[scale=0.5]}](11.5,2) to[bend left=40] (12.75,3);
        \draw[<->, red, thick,>={Latex[scale=0.5]}](16.5,2) to[bend left=40] (17.75,3);

        \draw[<->, red, thick,>={Latex[scale=0.5]}](1.5,6) to[bend left=40] (2.75,7);
        \draw[<->, red, thick,>={Latex[scale=0.5]}](6.5,6) to[bend left=40] (7.75,7);
        \draw[<->, red, thick,>={Latex[scale=0.5]}](11.5,6) to[bend left=40] (12.75,7);
        \draw[<->, red, thick,>={Latex[scale=0.5]}](16.5,6) to[bend left=40] (17.75,7);

        \draw[-latex, thick] (9.5,-1) -- (9.5,-4);
        \node[right,font=\small\itshape] at (10.5,-2.5) {Exchange};
        \draw[-latex, thick] (10.5,-4) -- (10.5,-1);
    \end{scope}

    \begin{scope}[yshift=-39cm]
        \draw (0,0) rectangle (2,8);
        \node at (1,1) {24};
        \node at (1,3) {16};
        \node at (1,5) {8};
        \node at (1,7) {0};

        \begin{scope}[xshift=2.5cm]
            \draw (0,0) rectangle (2,8);
            \node at (1,1) {25};
            \node at (1,3) {17};
            \node at (1,5) {9};
            \node at (1,7) {1};
        \end{scope}

        \begin{scope}[xshift=5cm]
            \draw (0,0) rectangle (2,8);
            \node at (1,1) {26};
            \node at (1,3) {18};
            \node at (1,5) {10};
            \node at (1,7) {2};
        \end{scope}

        \begin{scope}[xshift=7.5cm]
            \draw (0,0) rectangle (2,8);
            \node at (1,1) {27};
            \node at (1,3) {19};
            \node at (1,5) {11};
            \node at (1,7) {3};
        \end{scope}

        \begin{scope}[xshift=10cm]
            \draw (0,0) rectangle (2,8);
            \node at (1,1) {28};
            \node at (1,3) {20};
            \node at (1,5) {12};
            \node at (1,7) {4};
        \end{scope}

        \begin{scope}[xshift=12.5cm]
            \draw (0,0) rectangle (2,8);
            \node at (1,1) {29};
            \node at (1,3) {21};
            \node at (1,5) {13};
            \node at (1,7) {5};
        \end{scope}

        \begin{scope}[xshift=15cm]
            \draw (0,0) rectangle (2,8);
            \node at (1,1) {30};
            \node at (1,3) {22};
            \node at (1,5) {14};
            \node at (1,7) {6};
        \end{scope}

        \begin{scope}[xshift=17.5cm]
            \draw (0,0) rectangle (2,8);
            \node at (1,1) {31};
            \node at (1,3) {23};
            \node at (1,5) {15};
            \node at (1,7) {7};
        \end{scope}

        \node[left,font=\small\itshape] at (7,9.5) {Permuted:};
    \end{scope}

\end{tikzpicture}
    $}
\caption{Block permutation for $L = 4$, $N = 8$.  Each
column represents a block of $L$ samples.}
\label{fig:block_permutation}
\end{figure}

To guide the general description below,
Figure~\ref{fig:block_permutation} illustrates the block
permutation procedure for the case $L = 4$, $N = 8$.
While the block size~$L$ is determined by the native
hardware parallelism, the number of blocks~$N$ is a tuning parameter that
can be increased to reduce the per-sample operation,
as shown in the complexity counts of
Tables~\ref{tab:ph_complexity}
and~\ref{tab:cr_complexity}.
For efficient permutation, the number of blocks $N$ must be an integer multiple of the block size $L$.

The permutation proceeds in two phases.  
First, the $N$
blocks are regrouped by selecting every $(N/L)$-th block,
so that the $N$ blocks are partitioned into $N/L$ groups of
$L$ blocks each. 
In the second step, the $L$ blocks of each group are transposed
using
$\log\ssub{2} L$
rounds of anti-diagonal exchanges.
At each level,
elements in the lower-left quadrant of each sub-matrix are
swapped with the corresponding elements in the upper-right
quadrant, with the sub-matrix size halving at each level.
The de-permutation is obtained by applying the inverse
operations in reverse order.

The regrouping phase requires only block-wise memory access and no processing.  
Each of the $\log\ssub{2} L$ exchange levels
requires $N$ intra-block shuffles (one per block).
Permutation and de-permutation, each require
$N \log\ssub{2} L$ shuffles, or equivalently
$\log\ssub{2} L / L$ shuffles per sample.

\subsection{Realization of High-Order Systems}
\label{sec:HOS}

A general IIR filter of order~$M$ is realized as a cascade
of~$K=\lceil \frac{M}{2} \rceil$ biquads~\eqref{eq:general_cascaded_equation}, where the output of
section~$k$ serves as the input to section~$k{+}1$.  In
scalar filtering, each biquad costs~$4$ FMAs per sample; in
block filtering, each biquad costs~$1 + 4/L$ block FMAs per
sample.  In both cases, the cascade cost scales linearly
with~$K$.

For the proposed multi-block filtering algorithms, the
cascade admits an important simplification.  Each biquad
operates on the stride-$N$ permuted layout, and the output
of section~$k$ is already in the permuted form required by
section~$k{+}1$.  Consequently, the de-permutation at the
output of one section and the re-permutation at the input
of the next are inverse operations that cancel exactly.
Only a single permutation at the input and a single
de-permutation at the output of the entire cascade are
required.  The permutation cost of~$2\log\ssub{2} L / L$ shuffles
per sample is therefore amortized over the~$K$ biquads,
reducing to~$2\log\ssub{2} L / (KL)$ shuffles per sample per
biquad.

Table~\ref{tab:overall_complexity} compares the per-sample,
per-biquad costs of the filtering algorithms for the
cascaded realization.  
Block filtering reduces the
per-sample operation count from~$4$ in scalar filtering
to~$1 + 4/L$, reflecting the intra-block parallelism
where~$L$ samples within each block are computed
simultaneously.  
Cyclic Reduction as well as the LU and PH factorizations further
reduce the dominant per-sample cost to~$6/L$, reflecting
the inter-block parallelism gained by processing~$N$ blocks
jointly.  
PH improves upon LU in both block FMA count and
sequential depth, as it replaces the computation involving
dense matrices with recursive doubling; the trade-off is a
moderate increase in shuffle count.  
Cyclic reduction further
exploits inter-block parallelism, reducing the sequential
depth from~$\mathcal{O}(1/L)$ of PH to~$\mathcal{O}(\log\ssub{2} N / (NL))$ per
sample.\looseness=-1

\begin{table}[t]
\centering
\caption{Per-sample, per-biquad complexity for the cascaded
realization of a $2K$-th order IIR filter.}
\label{tab:overall_complexity}
\setlength{\tabcolsep}{2.8pt}
\scriptsize
\begin{tabular}{lccc}
\toprule
 & & & Per-sample \\
Algorithm & Block FMAs & Shuffles & seq.\ depth $\mathcal{O}(\cdot)$ \\[1pt]
\midrule
Scalar filtering
  & $4$ & --- & $1$ \\[2pt]
Block filtering
  & $1 + \frac{4}{L}$ & --- & $\frac{\log\ssub{2} L}{L}$ \\[4pt]
LU factorization
  & $\frac{6}{L} {+} \frac{2}{N}
    {+} \frac{\log\ssub{2} L{-}4}{NL}$
  & $\frac{2\log\ssub{2} L}{KL}
    {+} \frac{\log\ssub{2} L{+}4}{NL}$
  & $\frac{1}{L} {+} \frac{3\log\ssub{2} L}{NL}$ \\[4pt]
PH factorization
  & $\frac{6}{L}
    {+} \frac{2\log\ssub{2} L{-}4}{NL}$
  & $\frac{2\log\ssub{2} L}{KL}
    {+} \frac{4\log\ssub{2} L{+}5}{NL}$
  & $\frac{1}{L} {+} \frac{\log\ssub{2} L}{NL}$ \\[4pt]
Cyclic reduction
  & $\frac{6}{L} {+} \frac{1}{N} {-} \frac{1}{NL}$
  & $\frac{2\log\ssub{2} L}{KL}
    {+} \frac{2\log\ssub{2} N{+}4}{NL}$
  & $\frac{2\log\ssub{2} N{+}\log\ssub{2} L}{NL}$ \\
\bottomrule
\end{tabular}
\end{table}

\section{Experimental Results}
\label{sec:results}

In this section, we address the performance of our algorithms experimentally to support the analysis based on operation counts.
We begin by demonstrating that multi-block filtering does not sacrifice numerical accuracy relative to sequential filtering.
Then, fine-grained measurements based on hardware performance counters are used to measure performance of algorithm blocks and entire filters.

\subsection{Numerical Accuracy}
\label{sec:accuracy}

For recursive filters, numerical accuracy is of great concern as rounding errors have the potential to render filters unstable.
Cascaded biquads are used in practice to mitigate such problems and, hence, our experiment focuses on the numerical accuracy of second-order systems.

Two experiments are conducted: the first focuses on a
single biquad while the second considers tenth-order
elliptic bandpass and lowpass designs.
For both experiments, the input signal is a pseudo-random, white Gaussian noise signal with 2048 samples. 
The reference output signal is obtained from sequential scalar filtering with all computations performed using double-precision floats.
The accuracy metric is the normalized mean-squared error (MSE), computed as the power of the difference between the observed output and the reference output signals  divided by the power of the 
output signal.
For all algorithms, the derived parameters -- the block-filtering
matrix~$\Hb$~\eqref{eq:block_filtering_matrices}, the PH
coefficients of~\eqref{eq:U_12} and the powers
of~$\bm{C}$~\eqref{eq:C}, as well as the level
coefficients~\eqref{eq:cr_coeff_update} for cyclic
reduction --- are pre-computed 
in double precision and then rounded to single precision for filtering.


For the first experiment, a single biquad is considered with the angle $\theta$ of the pole as the parameter.
Specifically, the numerator coefficients are~$b\ssub{1} = -1.2$, $b\ssub{2} = 0.7$ and the poles are located at~$re^{\pm j\theta}$ with~$r = 0.9$ ($a\ssub{1} = -1.8\cos(\theta)$, $a\ssub{2} = 0.81$).
The block length is $L=8$ and $N=16$ blocks are used for the multi-block algorithms.

\begin{figure}[!t]
\centering
\includegraphics[width=\columnwidth]{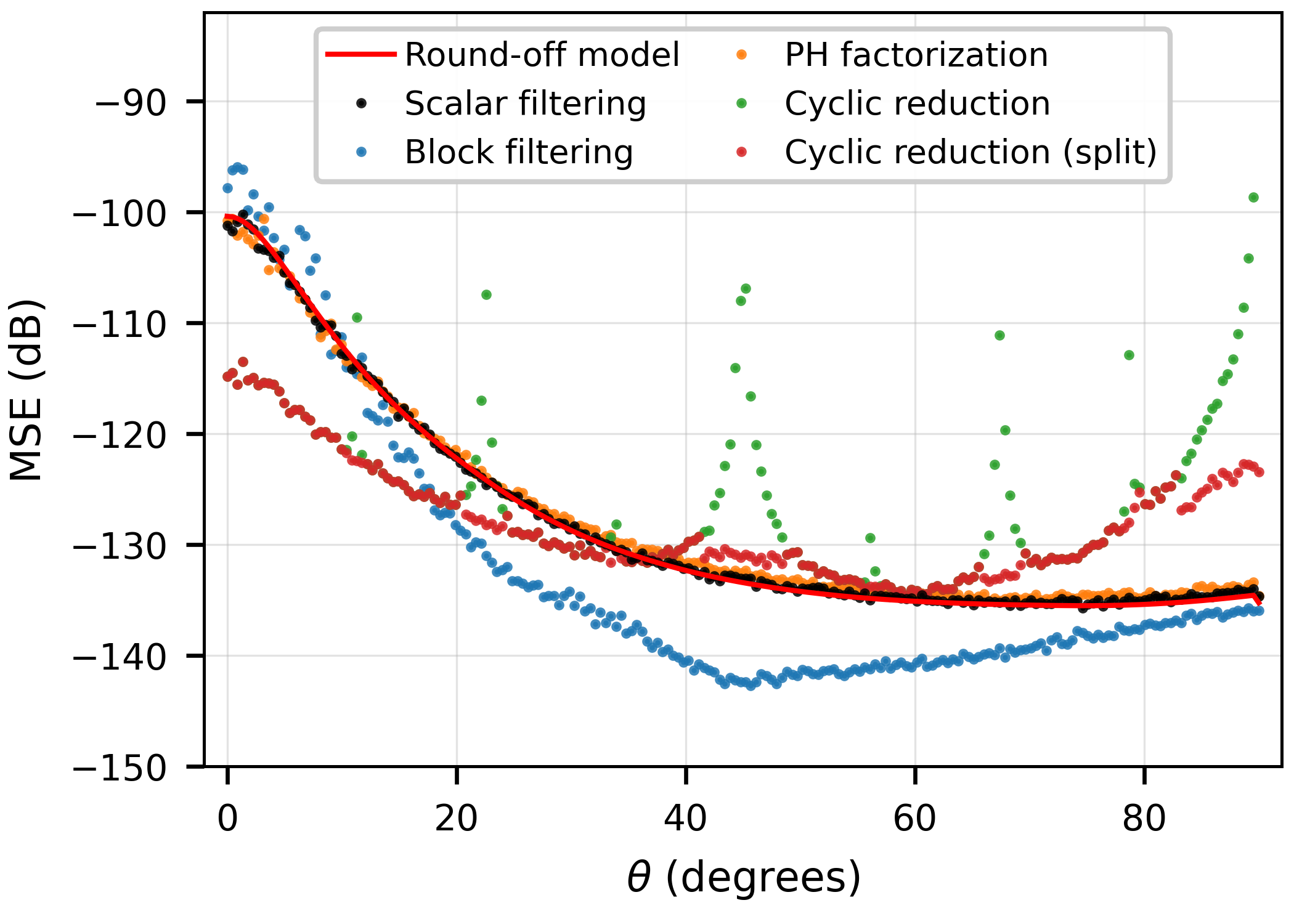}
\caption{Round-off error of one second-order section
versus the pole angle~$\theta$: single-precision MSE
against the double-precision reference, with the round-off
model~\eqref{eq:roundoff_model}.  The split is applied
only where triggered; elsewhere the split curve coincides
with cyclic reduction.  At~$\theta = 90^{\circ}$ the CR
error reaches~${+}183$\,dB, off scale.}
\label{fig:sos_accuracy}
\end{figure}


%
For scalar filtering~\eqref{eq:complex_second_order},
the classical round-off
analysis~\cite[Section 6.9]{Oppenheim_10}, \cite{Liu_69}
leads to the model
\begin{equation}
\label{eq:roundoff_model}
\sigma\ssub{f}^2 \approx
\underbrace{\sigma\ssub{\varepsilon}^2
\Bigl(\sum\nolimits_k P\ssub{k}\Bigr)}_{\text{injected noise power}}
\;
\underbrace{\frac{1+r^2}{1-r^2}\cdot
\frac{1}{1+r^4-2r^2\cos 2\theta}}_{\text{noise gain}},
\end{equation}
%
where $\sigma\ssub{\varepsilon}^2 \approx 0.18 \cdot 2^{-48}$ is
the variance of the relative rounding error of a
single-precision operation, $P\ssub{k}$ is the power of
the $k$-th rounded product or sum in the scalar evaluation
of~\eqref{eq:complex_second_order}, and the noise gain is
the energy of the impulse response of the all-pole part,
through which every injected rounding recirculates.  
Figure~\ref{fig:sos_accuracy} shows that the 
measured error for scalar filtering matches the model prediction.

Not surprisingly, PH factorization also matches this model, as the vectorized computations~\eqref{eq:feedforward_block_uniform}, \eqref{eq:particular_and_homogeneous}, and~\eqref{eq:PH_forward} correspond closely to their scalar counterparts. 

Block filtering achieves lower MSE for most pole locations, especially for poles near $45^{\circ}$.
This is the result of pre-computing segments of the filter's impulse response (in the matrix $\Hb = \Ab^{-1}\,\Bb$ in~\eqref{eq:block_filtering_matrices}) using double-precision before rounding.
As with scalar and PH filtering, MSE is largest for poles with $\theta \approx 0^{\circ}$ when two poles are near the real axis.

Cyclic reduction exhibits the most interesting behavior.
Without splitting (green dots in Figure~\ref{fig:sos_accuracy}), MSE spikes when the pole angle $\theta^{(i)}$ is near~$90^{\circ}$, i.e., $a\ssub{1}^{(i)}\approx 0$ in any reduction round;
this is the behavior predicted in Section~\ref{sec:cyclic_reduction}.
Figure~\ref{fig:sos_accuracy} also shows that the proposed remedy --- to apply splitting whenever
$|a\ssub{1}^{(i)}| < \frac{1}{4} |a\ssub{2}^{(i)}|$ --- is effective in eliminating these spikes.
Cyclic reduction has the smallest maximum MSE of all algorithms.
For poles near the real axis, the MSE is reduced since each reduction round moves the poles further away from the unit circle.
Poles near the imaginary axis show higher MSE than the other algorithms since reduction maps these poles to locations near the real axis.


\begin{table}[!t]
\caption{Normalized MSE (dB) of the tenth-order elliptic
bandpass at a quarter of the sampling rate and lowpass with passband edge $0.02 f\ssub{s}$ ($2048$ white-noise samples, $L = 8$,
$N = 16$).  The reference is the double-precision scalar
run.}
\label{tab:cascade_accuracy}
\centering
\begin{tabular}{lcc@{\hskip 1.2em}cc}
\toprule
 & \multicolumn{2}{c}{Bandpass} & \multicolumn{2}{c}{Lowpass} \\
\cmidrule(lr){2-3} \cmidrule(lr){4-5}
Algorithm & Double & Single & Double & Single \\
\midrule
Scalar filtering & --- & $-128$ & --- & $-94$ \\
Block filtering & $-298$ & $-130$ & $-259$ & $-89$ \\
PH factorization & $-300$ & $-128$ & $-256$ & $-88$ \\
Cyclic reduction & $-199$ & $-40$ & $-264$ & $-101$ \\
Cyclic reduction (split) & $-277$ & $-120$ & $-264$ & $-101$ \\
\bottomrule
\end{tabular}
\end{table}


The second experiment applies the filtering algorithms to
two complete tenth-order elliptic designs---five
second-order sections, $0.5$\,dB passband ripple,
$60$\,dB stopband attenuation.
The bandpass design, with passband from $0.24 f\ssub{s}$
to $0.26 f\ssub{s}$, places every pole pair
near~$\theta = 90^{\circ}$, so every section triggers the
split at level~$0$; a small offset keeps the
coefficients~$a\ssub{1}$ from being exactly zero, to
prevent division by zero in CR without splitting.  The
lowpass design, with passband edge $0.02 f\ssub{s}$,
places every pole pair near~$\theta = 0^{\circ}$, so no
section triggers the split, and plain and split cyclic
reduction coincide.
Table~\ref{tab:cascade_accuracy} reports the normalized
MSE under the protocol of Figure~\ref{fig:sos_accuracy}
for single and double precision.  On the bandpass, pure
cyclic reduction gives up roughly $100$\,dB in either
precision, while the other algorithms do not.  On the
lowpass, cyclic reduction attains the lowest
single-precision MSE of all algorithms.

\subsection{Throughput}
\label{sec:throughput}

The throughput experiments augment the operation counts presented earlier.
Fine grained measurements account for difficult to predict effects of modern microprocessor architectures, including deep execution pipelines, multiple execution units, cache effects, and different execution times for instructions.
The analysis aims to quantify the actual performance of our algorithms and to identify potential bottlenecks.

Four algorithms are evaluated: scalar
filtering,
block filtering, PH
factorization, and cyclic reduction, all implemented in C++
using CPU SIMD vector instructions.  
SIMD is a natural fit
for the block computation model, as each packed vector
instruction operates on~$L$ samples simultaneously.
Hardware intrinsics are accessed through the Vector
Class Library (VCL)~\cite{Agner_04}, which provides a
portable, high-level interface.  All numerical computations
use single-precision floating-point arithmetic.

The source code is compiled with the \texttt{-O2} optimization flag
and \texttt{-march=native}; all loops are manually
unrolled, which eliminates loop-control overhead and
guarantees consistent instruction-level parallelism
independent of compiler heuristics.

To assess generality, we evaluate on three Intel
micro-architectures spanning a decade: Haswell, the
earliest generation supporting AVX2
(\SI{3.3}{\giga\hertz}), Skylake
(\SI{4.0}{\giga\hertz}), and the hybrid Meteor Lake
(\SI{4.5}{\giga\hertz}), where all measurements run on
the performance cores.

To enable fair comparison across multiple micro-architectures,
throughput is reported in clock cycles per sample (CPS).
Measurements are obtained using \texttt{PMCTest} from Agner Fog's
software optimization toolbox~\cite{Agner_25}, which
supports a wide range of micro-architectures and relies on
detailed hardware performance counters.

\begin{figure}[t]
    \centering
    \includegraphics[width=0.46\textwidth]{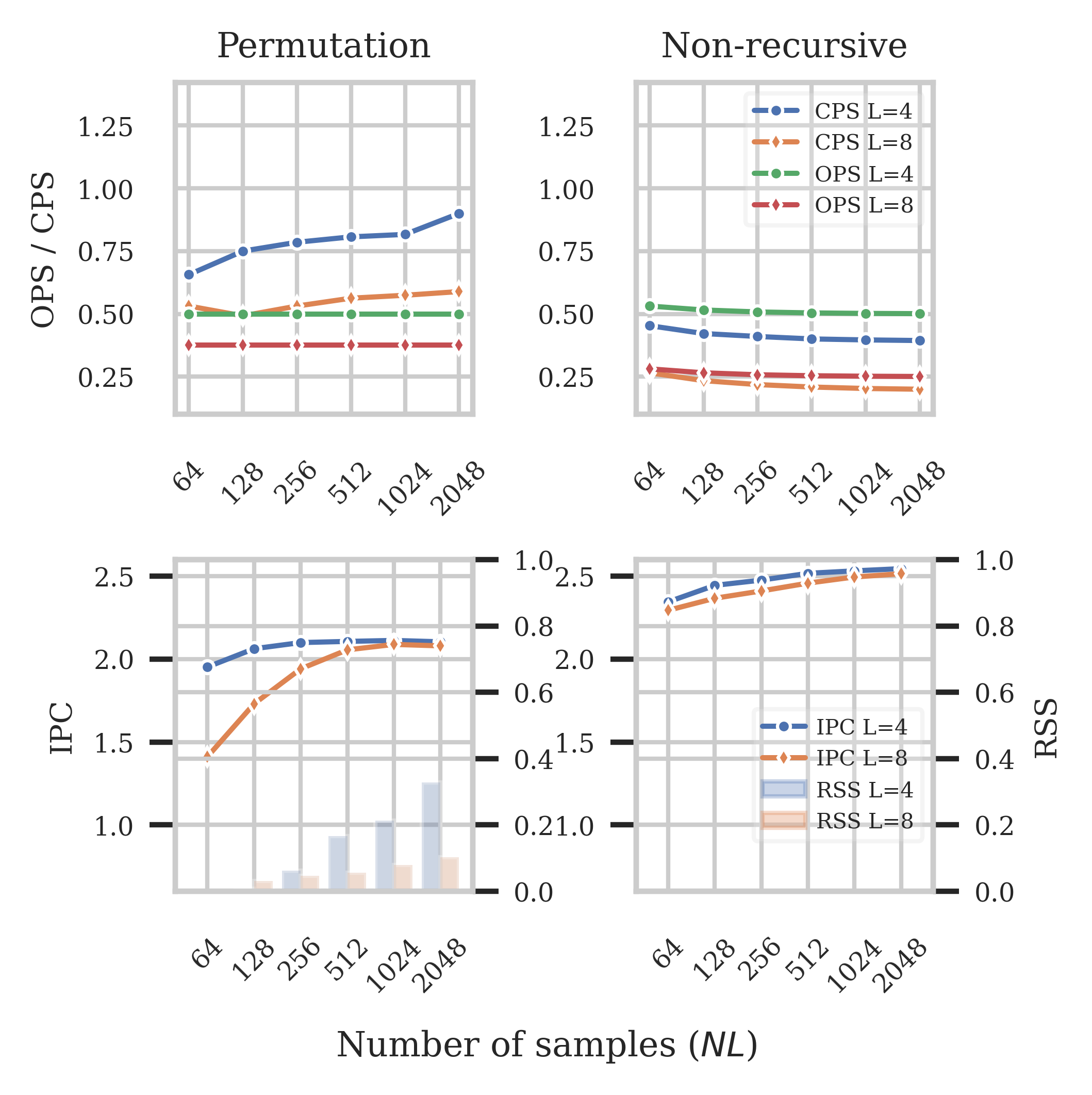}
    \caption{Measurements for common stages, permutation and non-recursive filtering, in multi-block filtering on Haswell versus sample size $NL$ for two block sizes $L$.}
    \label{fig:pm_fir}
\end{figure}

\begin{figure}[t]
    \centering
    \includegraphics[width=0.46\textwidth]{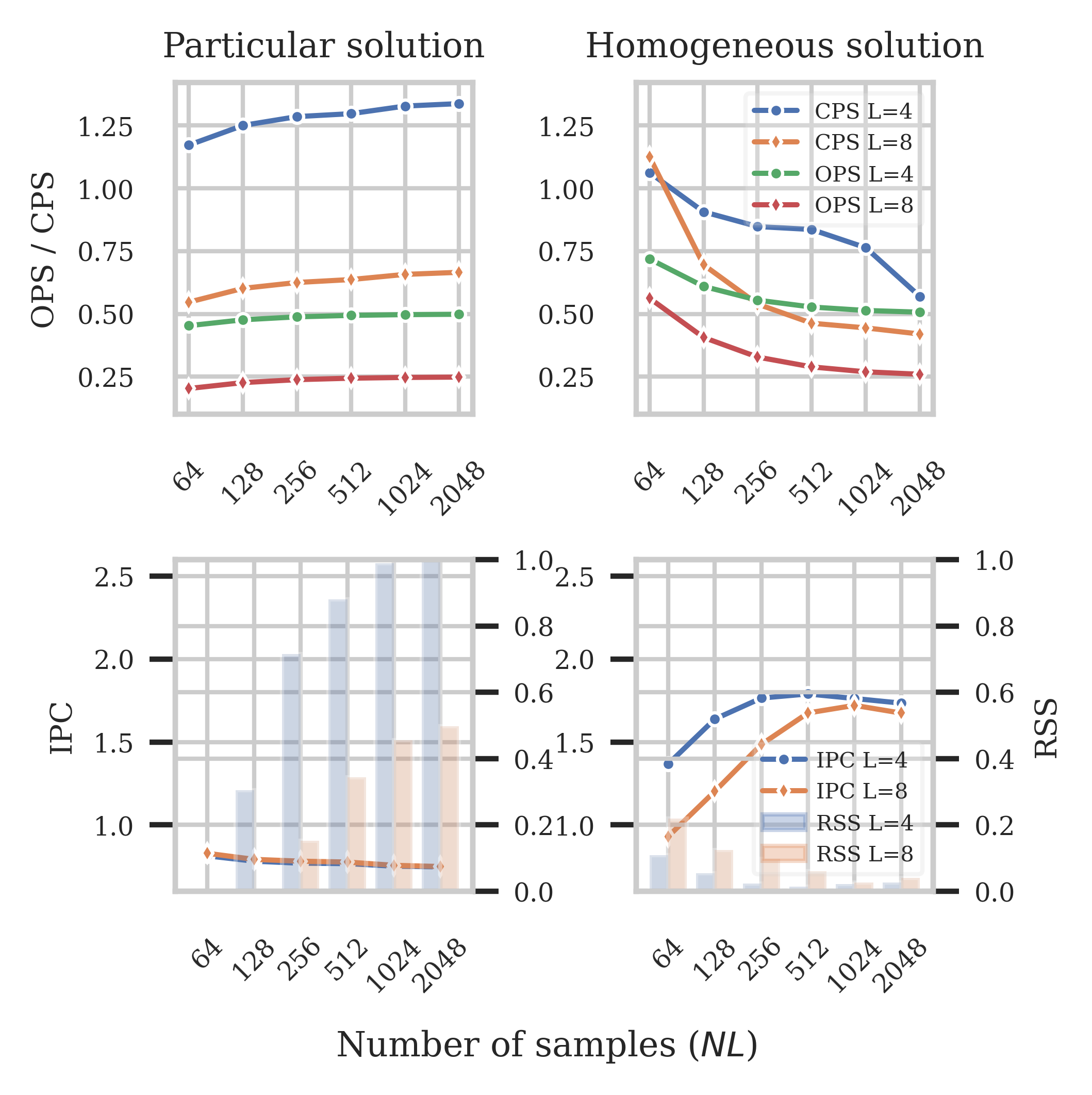}
    \caption{Measurements for PH factorization stages, particular and homogeneous solution, on Haswell versus sample size $NL$ for two block sizes $L$.}
    \label{fig:ph_22}
\end{figure}

\begin{figure}[t]
    \centering
    \includegraphics[width=0.46\textwidth]{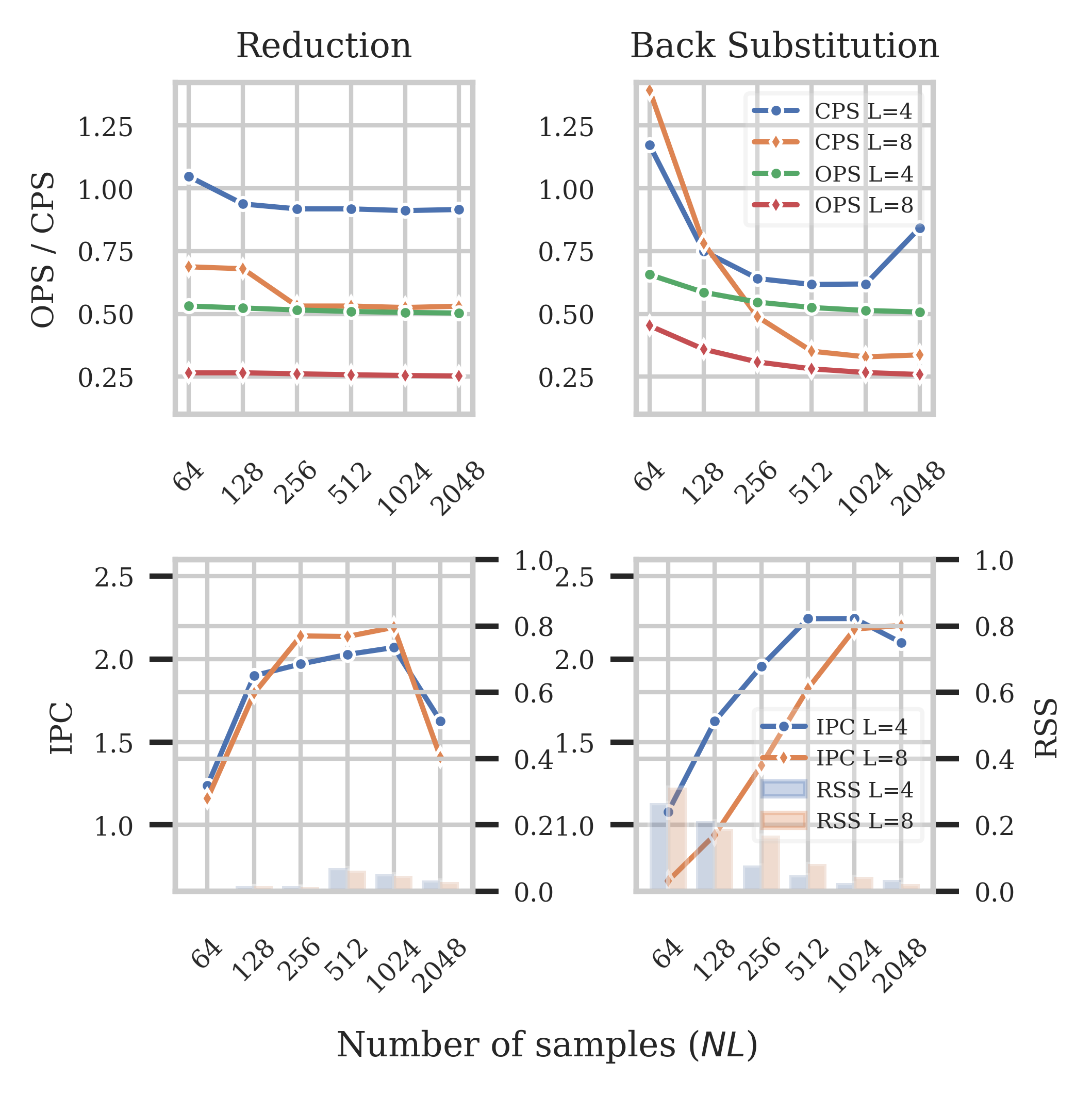}
    \caption{Measurements for Cyclic Reduction stages, reduction and back substitution, on Haswell versus sample size $NL$ for two block sizes $L$.}
    \label{fig:cr_22}
\end{figure}

Figures~\ref{fig:pm_fir}--\ref{fig:cr_22} present
the measured performance of the individual stages of
multi-block filtering on Haswell; 
the other architectures exhibit similar performance characteristics.
Throughout,
$L$ denotes the SIMD vector width and~$NL$ the number
of processed samples. Each figure consists of four panels.
The two top panels
compare measured clock cycles per sample (CPS) against the
nominal per-sample operation counts (OPS) derived from
Table~\ref{tab:ph_complexity} and Table~\ref{tab:cr_complexity},
which should be approximately proportional to each other. The two bottom
panels report micro-architectural behavior: instructions per
cycle (IPC) quantifies the attained instruction-level
parallelism, while resource-stall cycles per sample (RSS)
reflects how sequential dependencies limit pipeline concurrency.

Figure~\ref{fig:pm_fir} shows the permutation and
non-recursive stages. The CPS
for~$L=8$ is consistently lower than for~$L=4$,
confirming improved efficiency at wider SIMD widths.
For the permutation procedure, the intrinsic data dependencies 
require each \textit{exchange} step to complete
before the next can proceed. 
The non-recursive stage is
largely free of such dependencies, as all $N$~blocks can
be evaluated independently. Its CPS tracks the OPS curve
closely and improves with increasing~$N$ through improved
pipelining.


Figure~\ref{fig:ph_22} shows the particular and
homogeneous solution stages of PH factorization.
The particular solution and the non-recursive stage in
Figure~\ref{fig:pm_fir} exhibit similar OPS curves, yet
their CPS profiles differ substantially. 
This contrast
arises from the deep sequential dependency chain in the
particular solution.
The dependency prevents pipelining for the FMA operations so 
that each FMA incurs a 5-cycle latency (instead of 2~FMA per cycle with full pipelining).
The RSS plot
confirms the pipeline stalls: the particular solution exhibits
monotonically increasing resource stalls with
growing~$N$, accompanied by decreasing IPC.
The homogeneous solution, whose dominant data dependency
arises from the $\log\ssub{2} L$ levels of recursive
doubling rather than from the block count~$N$, shows
correspondingly lower stall cycles that decrease
with~$N$.

Figure~\ref{fig:cr_22} shows the reduction and
back-substitution stages of cyclic reduction, confirming
the theoretical advantage of $\mathcal{O}(\log\ssub{2} N)$
sequential depth over the~$\mathcal{O}(N)$ depth of the particular
solution. 
The RSS curves remain significantly lower than
those of the particular solution in
Figure~\ref{fig:ph_22}, particularly at large~$N$. 
The back substitution stage behaves analogously to the
homogeneous solution as the terminal block filtering
dominates at small~$N$.

\begin{figure}[t]
    \centering
    \includegraphics[width=0.46\textwidth]{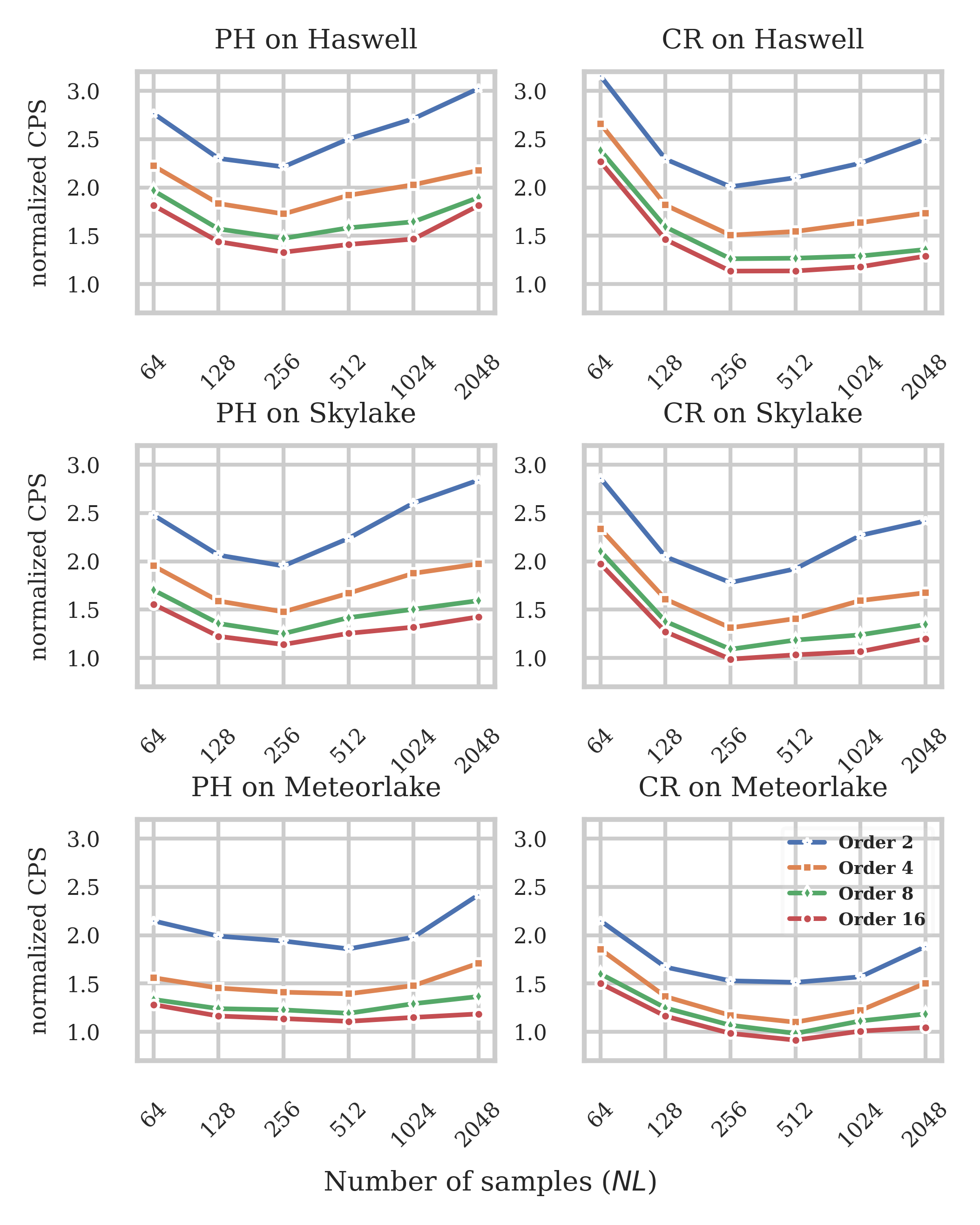}
    \caption{Normalized CPS of PH factorization and cyclic reduction filters across Haswell, Skylake, and Meteor Lake for orders 2, 4, 8, and 16 $(L = 8)$.}
    \label{fig:plot3}
\end{figure}

Figure~\ref{fig:plot3} compares CPS for PH factorization and cyclic reduction across different IIR system orders and micro-architectures.
Throughout, CPS is normalized by the number of cascade stages. 
The normalized CPS decreases as system order increases because the
permutation before and after filtering are amortized over more stages. 
A direct comparison between PH
factorization and cyclic reduction is most informative at
the highest tested order: PH achieves lower CPS at
small~$N$, where the $\mathcal{O}(N)$ sequential dependency of the
particular solution has not yet become the bottleneck.
Cyclic reduction attains its minimum normalized CPS at
larger~$N$ (signal block group size $NL = 512$, $N = 64$ on Meteor Lake),
where its $\mathcal{O}(\log\ssub{2} N)$ dependency
depth provides increasingly favorable pipeline
utilization. 
Both algorithms consistently achieve lower
CPS on newer architectures.

\begin{figure}[t]
    \centering
    \includegraphics[width=0.46\textwidth]{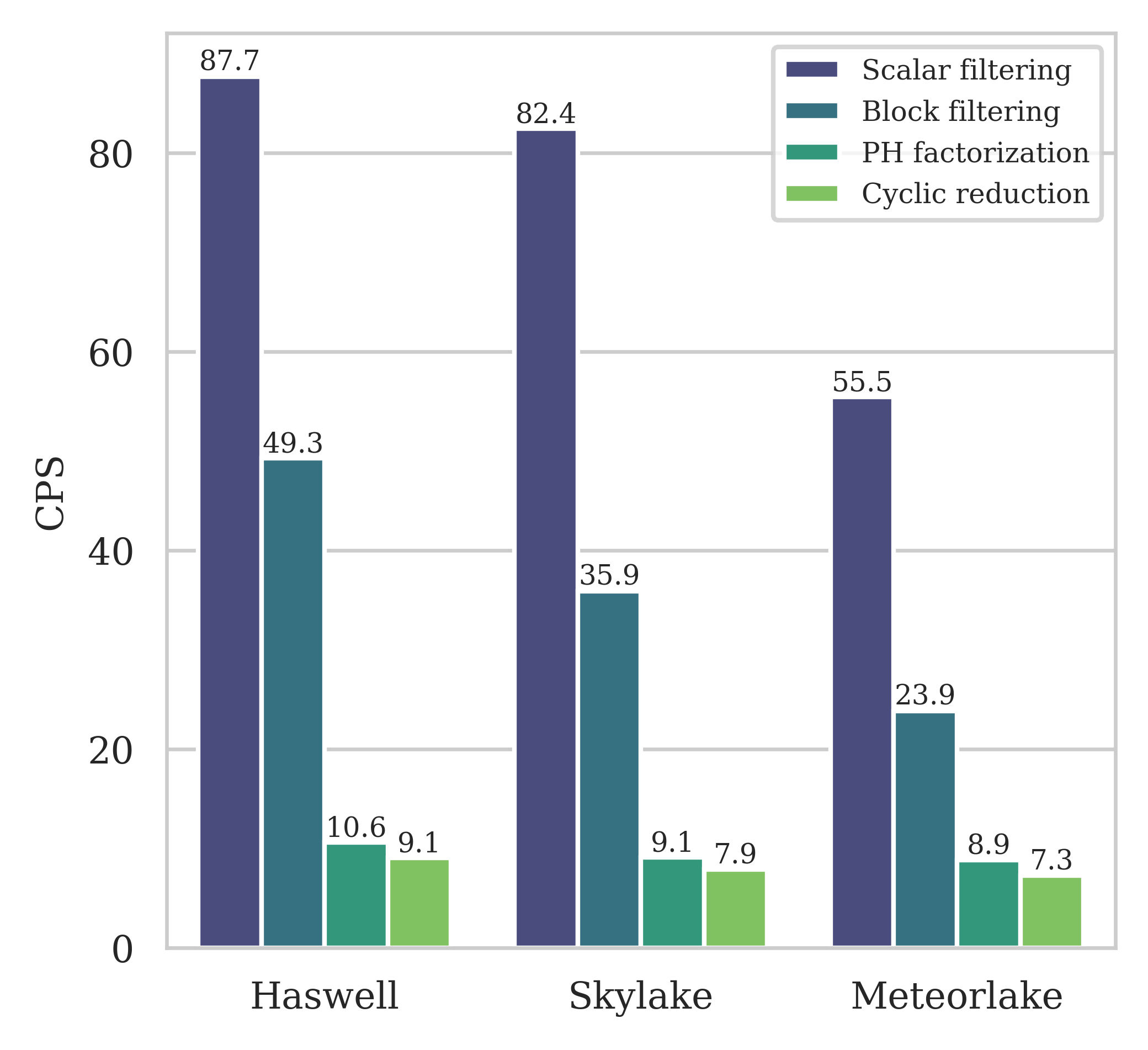}
    \caption{Best CPS comparison of four filtering algorithms for a 16th-order cascaded IIR filter across three architectures ($L=8$, $N$ chosen for highest throughput).}
    \label{fig:plot42}
\end{figure}

Figure~\ref{fig:plot42} compares the best CPS of scalar
filtering, block filtering, PH factorization, and cyclic
reduction for a 16th-order cascaded IIR filter. 
IIR filtering benefits substantially from vectorized
execution: block filtering reduces CPS by roughly
$2\times$ compared to the scalar baseline on Haswell, and
the two multi-block filtering algorithms reduce it
approximately $5 \times$ further still. 
Across all three micro-architectures, the
ordering is consistent: scalar filtering is the slowest,
followed by block filtering, PH factorization, and cyclic
reduction, confirming that the proposed algorithms scale
well with advances in processor design.


To place these results in the context of actual
throughput, consider a 16th-order cascaded IIR filter on
Meteor Lake at \SI{4.5}{\giga\hertz}.  Cyclic reduction
achieves the smallest CPS of $7.3$ (normalized CPS of $0.91 \times 8$ cascade stages)
at $NL = 512$, yielding a
throughput of $4.5 \times 10^{9} / 7.3 \approx 616$~MS/s.
For comparison, \texttt{scipy.signal.sosfilt}
\cite{sosfilt, virtanen_20}, which implements sequential
scalar filtering in~C, achieves approximately 77~MS/s
(58~CPS) on the same machine, consistent with the
55.5~CPS measured for scalar filtering in
Figure~\ref{fig:plot42}.
The proposed algorithm, thus,
delivers roughly an $8\times$ throughput improvement over
a standard sequential implementation on a single core.
Even Intel's vendor-tuned IPP biquad cascade
(\texttt{ippsIIR\_32f}, IPP~2026.0)~\cite{ipp_26}, which
reaches 151~MS/s on the same core by vectorizing the
numerators and pipelining the eight sections, remains
bound by the per-sample recurrence: cyclic reduction is
$4\times$ faster, the fastest implementation measured.

\section{Conclusion}
\label{sec:conclusion}


\noindent This paper presented a block-matrix
reformulation of cascaded second-order recursive
filtering that exposes parallelism at both the
intra-block and inter-block levels: cast as a banded
block-Toeplitz system and permuted with stride~$N$, the
recursive stage reduces to a block-tridiagonal structure
that admits efficient parallel solvers.  PH factorization
preserves the sparse block structure; cyclic reduction
halves the system at each level for
$\mathcal{O}(\log_2 N)$ sequential depth and, near a
quarter of the sampling rate, replaces the
ill-conditioned elimination by a division-free split.
The intermediate permutations between successive biquads
cancel exactly, amortizing the permutation overhead over
the entire cascade.

Experimental validation on three Intel micro-architectures
confirmed the theoretical complexity predictions.  For a
16th-order system, cyclic reduction achieves approximately
616~MS/s on a single Meteor Lake core, an $8\times$
throughput improvement over
\texttt{scipy.signal.sosfilt}, a $4\times$ improvement over Intel's
vendor-tuned IPP library, and up to $10\times$ reduction
in clock cycles per sample compared to scalar filtering.
Both algorithms scale favorably on newer architectures.
In single precision, block filtering, PH factorization,
and cyclic reduction (with splitting) all achieve numerical accuracy
comparable to scalar filtering.

All algorithms presented in this paper are released as an open-source C++ library, along with the raw measurement data, available at
\url{https://github.com/Haotian-RA/matrix_form_recursive_filtering}.
To the best of our knowledge, this is the first
open-source library that provides parallel IIR filtering
in  cascaded second-order form, combining good numerical
accuracy with very high throughput on commodity hardware.

A forthcoming companion paper \cite{zhai2026parallelcascadedrecursivefiltering} addresses the implementation of the
proposed algorithms on multi-core CPUs and GPUs, targeting
both high-throughput batched processing and low-latency
real-time scenarios.

\bibliographystyle{IEEEtran}

\bibliography{ref}



\end{document}